\documentclass[10pt,a4paper,twocolumn]{revtex4}

\usepackage[T1]{fontenc}
\usepackage[utf8]{inputenc}
\usepackage{lmodern}
\usepackage{amsmath}
\usepackage{array}
\usepackage{graphicx}

\usepackage{color}

\DeclareMathAlphabet{\mathwee}{OT1}{cmss}{m}{sl}

\newcommand{\hide}[1]{\relax}

\newcommand{\Og}{\ensuremath{\Omega}}
\newcommand{\Om}{\ensuremath{\Omega_\mathrm{m}}}
\newcommand{\OLC}{\ensuremath{\Omega_\mathrm{LC}}}

\newcommand{\Gm}{\ensuremath{\Gamma_\mathrm{m}}}
\newcommand{\GLC}{\ensuremath{\Gamma_\mathrm{LC}}}

\newcommand{\chim}{\chi_\text{m}}
\newcommand{\chiLC}{\chi_\text{LC}}

\newcommand{\bx}{\ensuremath{\bar x}}
\newcommand{\bq}{\ensuremath{\bar q}}

\newcommand{\mhat}{}

\newcommand{\kB}{\ensuremath{k_\mathrm{B}}}

\newcommand{\dhx}{\ensuremath{\delta \mhat x}}

\newcommand{\dhp}{\ensuremath{\delta \mhat p}}
\newcommand{\dhq}{\ensuremath{\delta \mhat q}}
\newcommand{\dhphi}{\ensuremath{\delta \mhat \phi}}

\newcommand{\dhFth}{\ensuremath{\delta\! \mhat F_\mathrm{th}}}

\newcommand{\dhVs}{\ensuremath{\delta\! \mhat V_\mathrm{s}}}

\newcommand{\dhV}{\ensuremath{\delta\! \mhat V}}

\newcommand{\dif}{\mathrm{d}}

\begin{document}
\title{Optical detection of radio waves through a nanomechanical transducer}
\author{T. Bagci$^{1}$, A. Simonsen$^{1}$, S. Schmid$^{2}$,  L. G. Villanueva$^{2}$, E. Zeuthen$^{1}$, J. Appel$^{1}$, J. M. Taylor$^{3}$,  A. S{\o}rensen$^{1}$, K. Usami$^{1}$, A. Schliesser$^{1,}$}
\email{albert.schliesser@nbi.dk}
\author{E. S. Polzik$^{1,}$}\email{polzik@nbi.dk}
\affiliation{$^{1}$Niels Bohr Institute, University of Copenhagen, Denmark}
\affiliation{$^{2}$Department of Micro--\ and Nanotechnology, Technical University of Denmark, DTU Nanotech, 2800 Kongens Lyngby, Denmark}
\affiliation{$^{3}$Joint Quantum Institute/NIST, College Park, Maryland, USA}

\begin{abstract}
\bf 
Low-loss transmission and sensitive recovery of weak radio-frequency (rf) and microwave signals is an ubiquitous technological challenge, crucial in fields as diverse as radio astronomy, medical imaging, navigation and communication, including those of quantum states.
Efficient upconversion of rf-signals to an optical carrier would allow transmitting them via optical fibers instead of copper wires dramatically reducing losses, and give access to the mature toolbox of quantum optical techniques, routinely enabling quantum-limited signal detection.
Research in the field of cavity optomechanics \cite{Kippenberg2008,Aspelmeyer2013} has shown that  nanomechanical oscillators can couple very strongly to either microwave \cite{OConnell2010, Teufel2011,Palomaki2013} or optical fields \cite{Groblacher2009a, Verhagen2011}.
An oscillator accommodating both these functionalities would bear great promise as the intermediate platform in a radio-to-optical transduction cascade.
Here, we demonstrate such an opto-electro-mechanical transducer following a recent proposal \cite{Taylor2011} utilizing a high-Q nanomembrane.
%
A moderate voltage bias ($V_\mathrm{dc}<10\,\mathrm{V}$) is sufficient to induce strong coupling \cite{Groblacher2009a,Verhagen2011,Teufel2011} between the voltage fluctuations in a radio-frequency resonance circuit and the membrane's displacement, which is \emph{simultaneously} coupled to light reflected off its metallized surface.
The circuit acts as an antenna; the voltage signals it induces are detected as an optical phase shift with quantum-limited sensitivity.
The corresponding half-wave voltage is in the microvolt range, orders of magnitude below that of standard optical modulators.
The noise added by the mechanical interface is suppressed by the electro-mechanical cooperativity $C_\mathrm{em}\approx 6800$ and has a temperature of $T_\mathrm{N}=T_\mathrm{m}/C_\mathrm{em}\approx 40\,\mathrm{mK}$, where $T_\mathrm{m}$ is the room temperature  at which the entire device is operated.
This corresponds to a sensitivity limit as low as $5\,\mathrm{pV/\sqrt{Hz}}$, or $-210\,\mathrm{dBm/Hz}$ in a narrow frequency band around $1\,\mathrm{MHz}$.
Our work introduces an entirely new approach to all-optical, ultralow-noise detection of classical electronic signals, and sets the stage for coherent upconversion of low-frequency quantum signals to the optical domain \cite{Taylor2011, Regal2011, Safavi2011, Wang2012a, McGee2013}.
 \end{abstract}

\maketitle

Opto- and electromechanical systems \cite{Kippenberg2008, Aspelmeyer2013} have gained considerable attention recently for their potential as hybrid transducers between otherwise incompatible (quantum) systems, such as photonic, electronic, and spin degrees of freedom \cite{Rabl2010, Stannigel2010, Safavi2011}.
Coupling of radio-frequency or microwave signals to optical fields via mechanics is particularly attractive for today's, and future quantum technologies. Photon-phonon transfer protocols viable all the way to the quantum regime have already been implemented in both radio- and optical-frequency domains separately \cite{Verhagen2011, Dong2012, Palomaki2013}.

\begin{figure}[htbp]
    \centerline{\includegraphics[width=\linewidth]{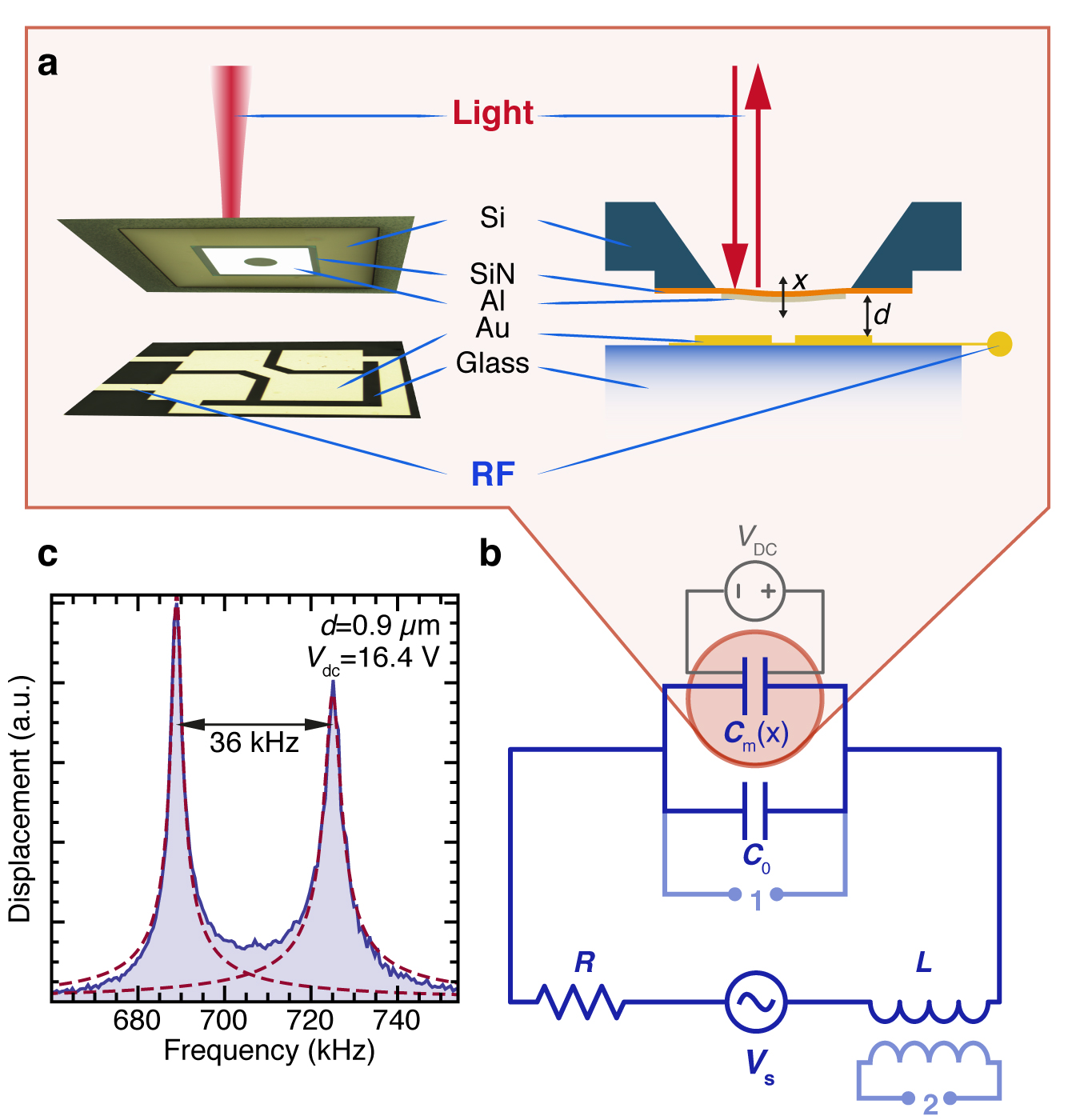}}
    \caption[exp scheme] {
    	\label{fig:setup} 
	\textbf{Optoelectromechanical system.} 
	The central part of the optoelectromechanical transducer ({\bf a}) is an Al-coated SiN $500 \mathrm{\mu m}$ square membrane in vacuum ($<10^{-5}$ mbar).
It forms  a position-dependant capacitor $C_\mathrm{m}(x=0)\approx 0.5\,\mathrm{pF}$ with a planar 4-segment  gold electrode in the immediate vicinity ($0.9\,\mathrm{\mu m} \lesssim d\lesssim 6\,\mathrm{\mu m}$). 
A laser beam is reflected off the membrane's Al coating \cite{Yu2012},  converting its displacement into a phase shift of the reflected beam.
({\bf b}) The membrane capacitor is part of a resonant $LC$-circuit,  tuned to the mechanical resonance frequency by means of a tuning capacitor $C_0\approx80\,\mathrm{pF}$ (see SI for details). 
A  bias voltage $V_\mathrm{dc}$ applied to the capacitor then couples the excitations of the $LC$-circuit to the membrane's motion.
The circuit is driven by a voltage $V_\mathrm{s}$ in series, which can be injected through the indicated coupling port `2' or picked up by the inductor from the ambient rf radiation.
({\bf c}) For a membrane-electrode distance of $0.9\mathrm{\mu m}$, the optically observed response of the membrane to a weak excitation of the system clearly shows a split peak (dashed lines: fitted Lorentzian resonances), due to hybridisation of the $LC$-circuit mode with the mechanical resonance of the membrane.
}
\end{figure}

Among the optomechanical systems that have been considered for radio-to-optical transduction \cite{Lee2010,Taylor2011, Safavi2011, McGee2013}, we choose an approach \cite{Taylor2011} based on a very high $Q_\mathrm{m} \approx 3 \cdot 10^5$ nanomembrane \cite{Thompson2007,Purdy2013} which is coupled capacitively \cite{Unterreithmeier2009,Schmid2010,Faust2012} to a radio-frequency (rf) resonance circuit, see Figure \ref{fig:setup}.
Together with a four-segment gold electrode, the membrane forms a capacitor, whose capacitance depends on the membrane-electrode distance $d+x$.
With a tuning capacitor $C_0$,  the total capacitance $C(x)=C_0+C_\mathrm{m}(x)$ forms a resonance circuit with a typical quality factor $Q_{LC}=\sqrt{L/C}/R$ of 130 when an inductor wired on a low-loss ferrite rod ($L=0.64\,\mathrm{mH}$) is used.
The inductor serves as an antenna feeding rf signals into the series $LC$-circuit.
The circuit's resonance frequency $\Omega_{LC}=1/\sqrt{LC}$ is tuned to the frequency $\Omega_\mathrm{m}/2 \pi =0.72\,\mathrm{MHz}$ of the fundamental drum mode of the membrane.
The  membrane-circuit system is  coupled to a propagating optical mode reflected off the membrane.

 The electromechanical dynamics is described most generically by the Hamiltonian \cite{Taylor2011} 
\begin{equation}
  \label{eq:ham}
  H=\frac{\phi^2}{2L}+\frac{p^2}{2m}+\frac{m\Omega_m^2x^2}{2}+\frac{q^2}{2C(x)}-qV_\mathrm{dc}
\end{equation}
where $\phi$ and $q$, the flux in the inductor and the charge on the capacitors, are conjugate variables for the $LC$ circuit; $x$ and $p$ denote the position and momentum of the membrane with an effective mass $m$.
The last two terms represent the charging energy $U_C(x)$ of the capacitors, which can be offset by an externally applied bias voltage $V_\mathrm{dc}$ (Fig.\ \ref{fig:setup}).
This energy corresponding to the charge $\bar q=V_\mathrm{dc} \, C(\bar x)$ leads to a new equilibrium position $\bar x$ of the membrane.
Furthermore, the position-dependent capacitive force $F_C(x)=-\frac{dU_C}{dx}$ causes  spring softening, reducing the membrane's motional eigenfrequency by  $\Delta \Omega_\mathrm{m} \approx -C''(\bar x) V_\mathrm{dc}^2/2 m \Omega_\mathrm{m}$ \cite{Unterreithmeier2009}.

Much richer dynamics than this shift may be expected from the mutually coupled system (\ref{eq:ham}).
For small excursions $(\delta q,\delta x)$ around the equilibrium $(\bar q, \bar x)$, it can be described by the linearised interaction term (Ref.\ \cite{Taylor2011} and SI)
\begin{equation}
  H_\mathrm{I}=G \delta q \delta x=\hbar g_\mathrm{em} \frac{ \delta q}{\sqrt{\hbar/2 L  \Omega_{LC}}} \frac{ \delta x}{\sqrt{\hbar/2 m \Omega_\mathrm{m}}},
\end{equation}
parametrized either by the coupling parameter $G=-V_\mathrm{dc}\, \frac{C'(\bar x)}{C(\bar x)}$ or the electromechanical coupling energy $\hbar g_\mathrm{em}$.
This coupling leads to an exchange of energy between the electronic and mechanical subsystems at the rate $g_\mathrm{em}$; if this rate  exceeds their dissipation rates $\Gamma_{LC}=\Omega_{LC}/Q_{LC}$, $\Gamma_\mathrm{m}=\Omega_\mathrm{m}/Q_\mathrm{m}$, they hybridise into a strongly coupled electromechanical system \cite{Groblacher2009a, Verhagen2011, Teufel2011}.
Our system is deeply in the strong coupling regime ($2g_\mathrm{em}=2\pi\cdot36\,\mathrm{kHz}>\Gamma_{LC} \gg\Gamma_\mathrm{m}$ for a distance $d=0{.}9\,\mathrm{\mu m}$ and a bias voltage of $V_\mathrm{dc}=16.4\,\mathrm{V}$ (Fig.\ \ref{fig:setup}c).
Here, for the first time, we detect the strong coupling  using an independent optical probe on the mechanical system.

We have performed an experimental series, in which the bias voltage is systematically increased, with a different sample and a larger distance $d=5{.}5\,\mathrm{\mu m}$.
The  system is excited inductively through port `2' (Fig.\ \ref{fig:setup}c), inducing a weak radio wave signal of (r.m.s) amplitude $V_\mathrm{s} =670\,\mathrm{nV}$, at a frequency $\Omega\approx\Omega_{LC}$.
The response of the coupled system can be measured both electrically as the voltage across the capacitors (port `1' in Fig.\ \ref{fig:setup}b) and optically
by analyzing the phase shift of a light beam (wavelength $\lambda=633\,\mathrm{nm}$) reflected off the membrane.
Both signals are recorded with a lock-in amplifier, which also provides the original excitation signal.

\begin{figure}[bp]
    \centerline{\includegraphics[width=\linewidth]{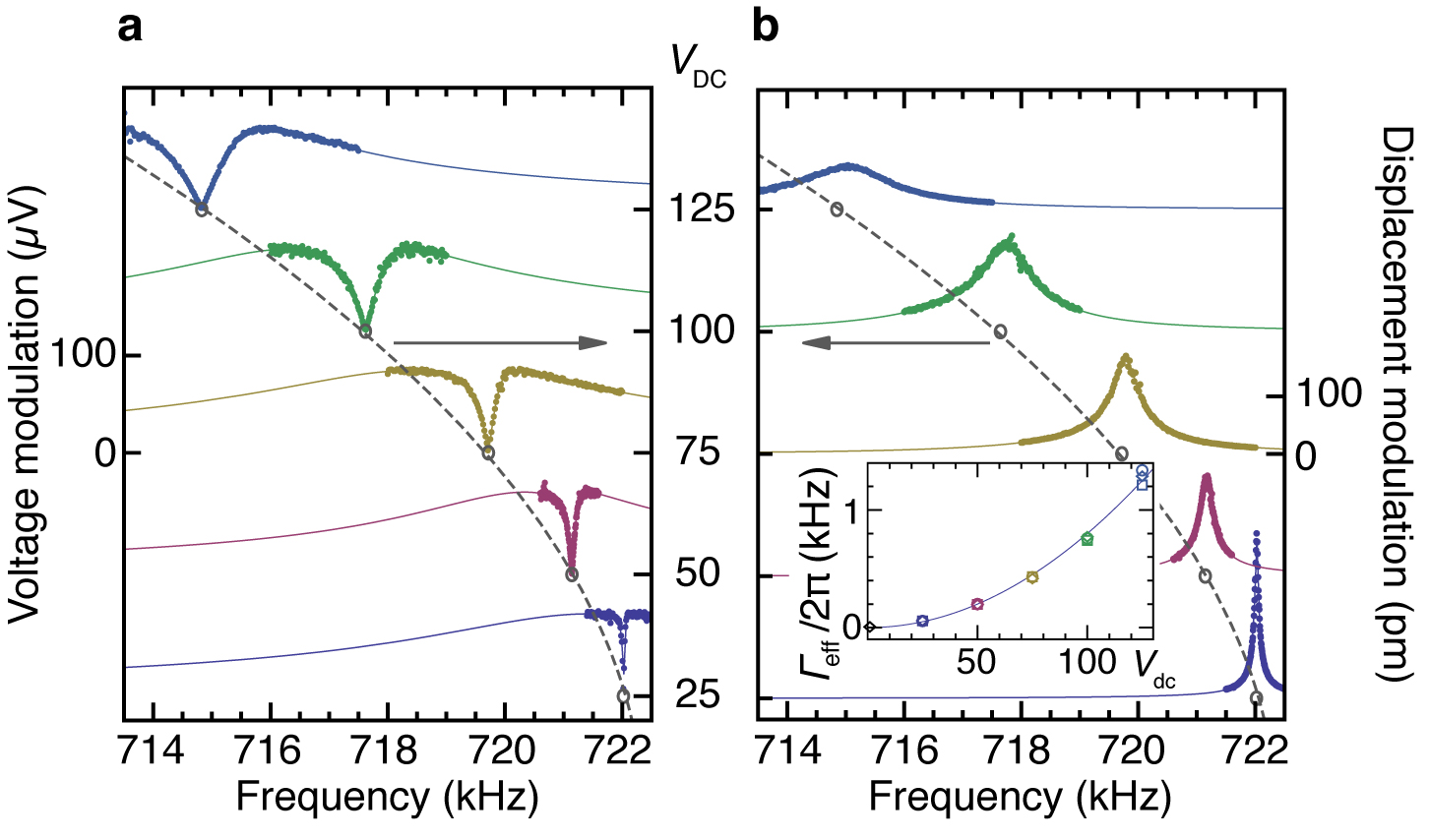}}
    \caption[mitscan] {
    	\label{fig:mit} 
	\textbf{Mechanically induced transparency.} 
	Response of the coupled system to weak excitation (through port `2' in Fig.\ \ref{fig:setup}b) probed though both, ({\bf a}) the voltage modulation in the $LC$ circuit (at port `1'), and ({\bf b}) the optical phase shift induced by membrane displacements.
The data (coloured dots) measured for five different bias voltages agree excellently with model fits (curves) corresponding to $g_\mathrm{em}/2\pi=\{ 280,470,810,1030,1290\} \, \mathrm{Hz}$.
Note that each curve is offset so that its baseline corresponds to the applied bias indicated on the scale between the panels.
Grey circles indicate the mechanical resonance frequency extracted for each set of data.
A  shift $\Delta \Omega_\mathrm{m}\propto -V_\mathrm{DC}^2$ can clearly be discerned (dashed line is a fit).
The inset shows the  effective linewidth of the mechanical resonance extracted from  full model fits to the electrically (circles) and optically (boxes) measured response and  simple Lorentzian fits to the optical data (diamonds). 
}
\end{figure} 

The electrically measured response (Fig.\ \ref{fig:mit}a) clearly shows the signature of a mechanically induced transparency \cite{Schliesser2010, Weis2010, 
Zhou2013} indicated by the dip in the  $LC$ resonance curve.
Independently, we observe the rf signal generated in the $LC$ circuit optically via the membrane mechanical dynamics  (Fig.\ \ref{fig:mit}b).
In particular, the electromechanical coupling  leads to  broadening of the mechanical resonance, an electromechanical damping effect analogous to optomechanical dynamical backaction cooling \cite{Gigan2006, Arcizet2006a, Schliesser2006}, to a new effective linewidth $\Gamma_\mathrm{eff}=(1+\mathcal{C}_\mathrm{em})\cdot \Gamma_\mathrm{m}$,
where $\mathcal{C}_\mathrm{em}$ is the electromechanical cooperativity 
\begin{equation}
\mathcal{C}_\mathrm{em}=\frac{4 g_\mathrm{em}^2}{\Gamma_\mathrm{m} \Gamma_{LC}}.
\end{equation}
The width of the induced transparency dip and the mechanical linewidth grow in unison, and in agreement with our expectations as $\Gamma_\mathrm{eff}\propto V_\mathrm{dc}^2$ (inset).
Both these features also shift to lower frequencies as the bias voltage is increased, following the expected $\Delta \Omega_\mathrm{m}\propto-V_\mathrm{dc}^2$ dependence \cite{Unterreithmeier2009}.
Note that in each experiment we have tuned the $LC$ resonance frequency to $\Omega_\mathrm{m}$. 

Using the model based on the full Langevin equations (SI), derived from the Hamiltonian (\ref{eq:ham}), we fit the electronically and optically measured curves, and obtain fit parameters $\Omega_\mathrm{m}$, $\Omega_\mathrm{LC}$, $\Gamma_{LC}$, and $G$ which for the two curves agree typically within $1\%$.
Together with 
the intrinsic damping $\Gamma_\mathrm{m}/2\pi=2.3\,\mathrm{Hz}$ determined independently from thermally driven spectra, the system's dynamics can be quantitatively predicted.
Our data analysis allows us to quantify the coupling strength in three independent ways, by (i) analysis of the mechanical responses' spectral shape, (ii) comparison of the voltage and displacement modulation amplitudes, and (iii) the frequency shift \cite{Unterreithmeier2009} of the mechanical mode, and compare these experimental values with (iv) a theoretical estimate taking the geometry of the electromechanical transducer into account (SI).
For $V_\mathrm{dc}=125~\mathrm{V}$ we find $G=10.3\,\mathrm{kV/m}$ following to the first method, and similar values using the three others (cf.\ SI), corroborating our thorough understanding of the system.

\begin{figure}[htbf]
    \centerline{\includegraphics[width=\linewidth]{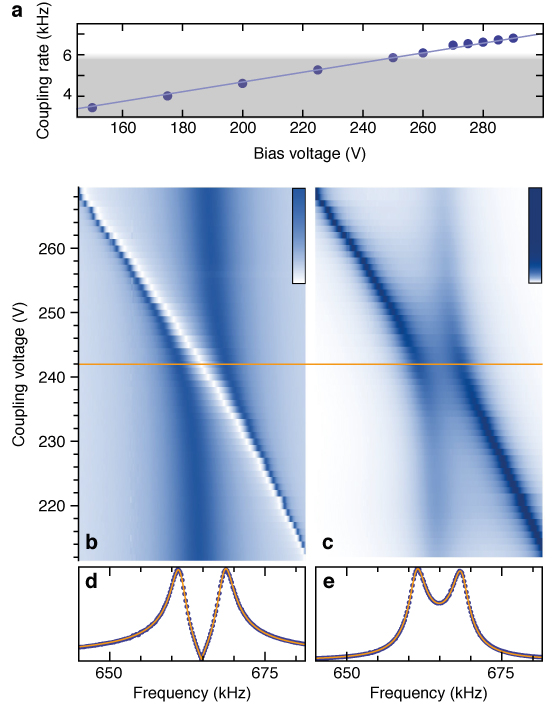}}
    \caption[strongcoupling] {
    		\label{fig:strong} \textbf{Strong coupling regime.} 
		({\bf a}) Measured coherent coupling rate $2 g_\mathrm{em}/2 \pi$ as a function of bias voltage (points) and linear fit (line).
The shaded area indicates the dissipation rate  $\Gamma_\mathrm{LC}/2 \pi\approx5.9\,\mathrm{kHz}$ of the LC circuit.
({\bf b}-{\bf e}) Normalised response of the coupled system as measured on port `1' (Fig.\ \ref{fig:setup}c) ({\bf b,d}) and via the optical phase shift induced by  membrane displacements ({\bf c,e}).
Upon tuning of the bias voltage the mechanical resonance frequency is tuned through the $LC$ resonance, but due to the strong coupling an avoided crossing is very clearly observed.
Panels ({\bf d,e}) show the spectra corresponding to the orange line in ({\bf b,c}), at $V_\mathrm{DC}=242\,\mathrm{V}$, where the electronic and mechanical resonance frequencies coincide.
Points are data, orange line is the fit of the model.
		}
\end{figure}

In another experimental run ($d=4{.}5\,\mathrm{\mu m}$, Fig.\ \ref{fig:strong}), we have characterised the strong electromechanical  coupling \cite{OConnell2010, Teufel2011,Palomaki2013} via the normal mode splitting \cite{Marquardt2007,Dobrindt2008} giving rise to an avoided crossing of the resonances of the electronic circuit and the mechanical mode, as the latter is tuned through the former using the capacitive spring effect \cite{Unterreithmeier2009}.
In contrast to earlier observations \cite{Groblacher2009a,Verhagen2011,Teufel2011} we can simultaneously witness the strong coupling through the optical readout, in which the recorded light phase reproduces the membrane motion (Fig.\ \ref{fig:strong}c,e).
Again, the predictions derived from the Langevin equations are in excellent agreement with our observations, yielding a cooperativity of $\mathcal{C}_\mathrm{em}=3800$ for these data with $m=24\,\mathrm{ng}$, $\Gamma_\mathrm{m}/2\pi = 3.1\,\mathrm{Hz}$.

We now turn to the performance of this interface as an rf-to-optical transducer.
A relevant figure of merit for the purpose of bringing small signals onto an optical carrier is the voltage $V_\pi$ required at the input of the series circuit in order  to induce an optical phase shift of $\pi$.
Achieving minimal $V_\pi$ requires a tradeoff between strong coupling and induced mechanical damping. 
For the optimal cooperativity $\mathcal{C}_\mathrm{em}=1$ we find 
\begin{equation}
  V_\pi=\frac{1}{2}\sqrt{m L \Gamma_\mathrm{m} \Gamma_{LC} } \lambda \Omega_\mathrm{r} \approx 140\,\mathrm{\mu V},
\end{equation}
at resonance $(\Omega_\mathrm{r}\equiv\Omega_\mathrm{m}=\Omega_{LC}=\Omega)$,
orders of magnitude below commercial modulators optimised for decades by the telecom industry, but also explorative microwave photonic devices \cite{Ilchenko2008b, Devgan2010, Hosseinzadeh2011} based on electronic nonlinearities \cite{Tsang2010}.
It is interesting to relate this performance to  more fundamental entities, namely the electromagnetic field's quanta that constitute the signal.
Indeed it is possible to show that the \emph{quantum} conversion efficiency, defined here as the ratio of optical sideband photons to the rf quanta extracted from the source $V_\mathrm{s} I /\hbar \Omega_\mathrm{LC}$ for $C_\mathrm{em}\gg1$, is given by (see SI) 
\begin{equation}
 \eta_\mathrm{eo} = 4(k x_\mathrm{zpf})^2 \frac{\Phi_\mathrm{car}}{\Gamma_\mathrm{m}}.
\end{equation}
This corresponds to the squared effective Lamb-Dicke parameter $ (k x_\mathrm{zpf})^2=(2 \pi/\lambda)^2 \hbar/(2 m \Omega_\mathrm{m})$ enhanced by the number of photons sampling the membrane during the membrane excitations' lifetime.
We have tested that the membranes can support optical readout powers of more than $\Phi_\mathrm{car}h c/\lambda=20\,\mathrm{mW}$ without degradation of their (intrinsic) linewidth. We thus project that conversion efficiencies on the order of $50\%$ are available.
So far we have measured a few percent in the laboratory, with further experiments ongoing.
Note that this transducer constitutes a phase-insensitive amplifier, and can thus reach conversion efficiencies above one---at the expense of added quantum noise.

For the recovery of classical signals, the sensitivity and bandwidth of the interface is of greatest interest.
The signal at  the optical output of the device is the interferometrically measured spectral density of the optical phase $\varphi$ of the light reflected off the membrane,
\begin{equation}
  S_{\varphi\varphi}^\mathrm{tot}=  (2k)^2\left| \chi_\mathrm{m}^\mathrm{eff} \right |^2 \left(
  				\left| G \chi_{LC} \right |^2 S_{VV}+S_{FF}^\mathrm{th} \right)+S_{\varphi\varphi}^\mathrm{im},
\end{equation}
The voltage $V_s$ at the input of the resonance circuit (denoted here as its spectral density $S_{VV}$) is transduced to a phase shift via the circuit's susceptibility $ \chi_{LC} $, the coupling $G$, the effective membrane susceptibility $ \chi_\mathrm{m}^\mathrm{eff}$ and the optical wavenumber $k$ (see SI).
The sensitivity is determined by the noise added within the interface.
This includes in particular, the imprecision in the phase measurement ($S_{\varphi\varphi}^\mathrm{im}$), but also the random thermal motion of the membrane induced by the Langevin force ($S_{FF}^\mathrm{th}$).
The former depends on the performance of the employed interferometric detector and can be quantum-limited ($S_{\varphi\varphi}^\mathrm{im}\sim \Phi_\mathrm{car}^{-1}$).

\begin{figure}[tbhp]
    \centerline{\includegraphics[width=.8\linewidth]{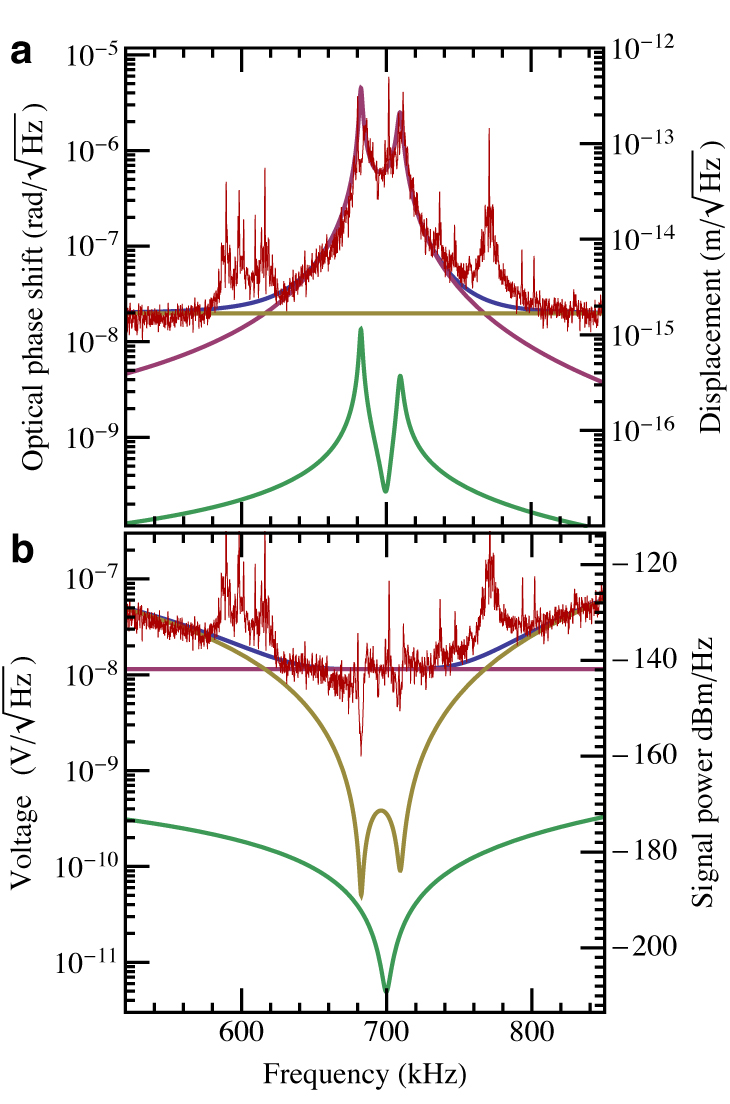}}
    \caption[sensitivitynoise] {
    \label{fig:sens} 
    {\bf Voltage sensitivity and noise.} 
    ({\bf a}) Optical measurent of  ambient rf radiation (red)  with a signal-to-noise ratio of nearly $50\, \mathrm{dB}$.  The sharp peak at $771\ \mathrm{kHz}$ is due to calibration with a known phase modulation. 
    Thick lines show models for the total signal ($\sqrt{S_{\varphi\varphi}^\mathrm{tot}}$, blue), with contributions from the rf radiation ($\sqrt{S_{VV}}$, violet), the optical quantum phase noise ($\sqrt{S_{\varphi\varphi}^\mathrm{im}}$, yellow), and membrane thermal noise ($\sqrt{S_{FF}^\mathrm{th}}$, green). 
    ({\bf b}) Data and models as in (a), but divided by the interface's response function $|\chi_\mathrm{tot}|$, and thus referenced to the voltage $V_\mathrm{s}$ induced in the antenna.
     }
\end{figure}

We demonstrate this transduction scheme by measuring the ambient 
 rf radiation background \cite{Kraus1966}.
This radiation induces a voltage $V_s$ on the order of $10\,\mathrm{nV}/\sqrt{\mathrm{Hz}}$ in the circuit, as we can determine through an electrical measurement on port `1'.
Alternatively, we measure this signal optically, as shown in Fig.\ \ref{fig:sens}a. 

In this experiment, we use a home-built interferometer operated at $\lambda=1064\,\mathrm{nm}$ and with a light power of $0.8\,\mathrm{mW}$ returned from a thicker membrane ($d\sim 1\,\mathrm{\mu m}$, $m=140\,\mathrm{ng}$, $\Gamma_\mathrm{m}/2\pi=20\,\mathrm{Hz}$).
Optical quantum (shot) noise limits the phase sensitivity to $\sqrt{S_{\varphi\varphi}^\mathrm{im}}=20\,\mathrm{nrad}/\mathrm{Hz}$, corresponding to membrane displacements of $1{.}7\,\mathrm{fm/\sqrt{Hz}}$.
This can be translated to a voltage sensitivity limit by division with the total input-output transfer function $\chi^\mathrm{tot}\equiv2k \chi_\mathrm{m}^\mathrm{eff}  G \chi_{LC}$ of the transducer.
Values as low as $50\,\mathrm{pV/\sqrt{Hz}}$ are achieved within the resonant bandwidth of this proof-of-principle transducer (yellow curve in Fig.\ \ref{fig:sens}b).
With improved detection (probing power $20\,\mathrm{mW}$, unity visibility), this number reduces to $10\,\mathrm{pV/\sqrt{Hz}}$.
Evidently, yet more sensitive optomechanical transduction could readily be achieved with known systems  \cite{Thompson2007,Anetsberger2009}.
The data in Fig.\ \ref{fig:sens} corresponds to a measurement with about $46\,\mathrm{dB}$ dynamic range, and  we still expect a significant margin towards the onset of mechanical nonlinearities, which were observed to occur only as the displacement amplitudes are approaching nm-levels, corresponding to an expected dynamic range of $80\,\mathrm{dB}$ for broadband signals.

The noise added by the membrane thermal agitation is more critical in nature, as it can usually only be reduced using cryogenic cooling.
Remarkably, however, this noise is strongly suppressed in this setting.
If  recast into an equivalent input voltage noise $S_{VV}^\mathrm{mem}(\Omega)=S_{FF}^\mathrm{th}(\Omega)/\left| G \chi_{LC}(\Omega) \right |^2$, this contribution is found as low as 
\begin{equation}
  S_{VV}^\mathrm{mem}(\Omega_{LC})=2 k_\mathrm{B} \frac{T_\mathrm{m}}{\mathcal{C}_\mathrm{em}}R,
\end{equation}
at resonance,  where $T_\mathrm{m}$ is the temperature of the \emph{membrane}.
Thus the noise added by the membrane, as seen by the optical readout, has a temperature of only $T_\mathrm{m}/\mathcal{C}_\mathrm{em}$.
For the data of Fig.\ \ref{fig:sens}, we have obtained a cooperativity of $\mathcal{C}_\mathrm{em}=6800$ with a bias voltage of $V_\mathrm{dc}=21\,\mathrm{V}$.
The corresponding noise temperature of $40\,\mathrm{mK}$ should therefore allow to recover signals down to a level of $5\,\mathrm{pV/\sqrt{Hz}}$ over a bandwidth of $\sim \Gamma_{LC}$ (green curve in Fig.\ \ref{fig:sens}b).

For comparison, the ultralow-noise operational amplifier arrangement, which we use to measure on port `1' has a nominal voltage noise of $4\,\mathrm{nV/\sqrt{Hz}}$ and negligible current noise; in practice, with a gain of $1000$, we achieve an input noise (at port `1') of $\sqrt{S_{VV}^\mathrm{oa}}=17\,\mathrm{nV/\sqrt{Hz}}$.
Referenced again to the voltage $V_\mathrm{s}$ driving the circuit, this translates to a voltage sensitivity of $\sqrt{S_{VV}^\mathrm{oa}}/Q_{LC}=130\,\mathrm{pV/\sqrt{Hz}}$, to be compared with the current shot-noise limited $50\,\mathrm{pV/\sqrt{Hz}}$ of our device and its $5\,\mathrm{pV/\sqrt{Hz}}$ membrane noise.
This superior performance of our transducer remains unchanged also off resonance ($\Omega\neq\Omega_{LC}$) as both methods benefit equally from the $LC$-circuit's resonant enhancement of the signal voltage $V_\mathrm{s}$ (cf.\ Fig. \ref{fig:setup}b).

Further improved sensitivity is readily available, not least due to the still considerable margin for increasing the cooperativity: the pull-in instability of the membrane is estimated to occur only at $\mathcal{C}_\mathrm{em}\approx 20{,}000$, and $S_{VV}^\mathrm{mem}(\Omega_\mathrm{R})\approx2.9\,\mathrm{pV/\sqrt{Hz}} $  in the present system (see SI).
Within the stability regime, stronger coupling, and correspondingly higher cooperativities, can be achieved by reducing the offset capacitance $C_0$.

The ultimate noise floor of our opto-mechanical transducer is well below the room temperature Johnson noise from the circuit's $R=20\,\mathrm{\Omega}$.
Our ultralow-noise transducer/amplifier can therefore be of particular relevance in applications where this noise is suppressed.
For example, for direct electronic (quantum) signal transduction, the resonance circuit must be overloaded with a cold transmission line which carries the signal of  interest. 
In radio astronomy  \cite{Kraus1966}, highly efficient antennas looking at the cold sky have noise temperatures in the GHz range significantly below room temperature.
The usually required cryogenically cooled pre-amplifiers might be replaced by our transducer.
Finally, in nuclear-magnetic resonance experiments including imaging, cooled pickup circuits can deliver a significant sensitivity improvement, yet this approach is challenging current amplifier technology \cite{Kovacs2005}.
For applications with the centre radio-frequency in the GHz band strong electro-mechanical coupling to the membrane in the MHz range can be achieved by using an oscillating coupling voltage \cite{Teufel2011, Palomaki2013} instead of the dc voltage used in the present work.

\vspace{0.5cm}
{\bf Methods Summary}
\vspace{0.5cm}

The capacitor is fabricated by standard cleanroom microfabrication techniques.
Electrodes made of gold (thickness of 200nm) are deposited on a glass substrate and structured by ion-beam etching.
Each segment is $400~\mu$m long, with $60~\mu$m gaps between the segments.
Pillars of a certain height (600~nm, $1\mu$m) are placed to determine the membrane-electrode distance.
The capacitor is connected in parallel to a ferrite  inductor with a $Q\approx500$ (around 700 kHz) and $L\approx635 \mu$H.
The inductor is wound with Litz wires to ensure high $Q$-factor.
A variable trimming capacitor is used to tune the resonance frequency of the LC circuit. The capacitance of the membrane-electrode system is measured to be roughly $0.5$pF.

The mechanical resonator consists of a 50~nm thick Aluminum layer on top of a  high-stress stoichometric SiN layer with a thickness of 100~nm and 180~nm for different samples.
The Al layer is deposited on top of the whole wafer after the membranes have been released.
Photolithography and chemical etching are subsequently used to remove the metal from the anchoring regions and from a circle in the middle of the membrane.
 The metal layer on SiN causes roughly a 10 percent decrease in the eigenfrequency of the fundamental mode.\

Optical interferometry is carried out via a commercial Doppler vibrometer (MSA-500 Polytec) and by a home made Michelson interferometer (for the data set in Fig.4). 
The vibrometer uses  optical heterodyne detection of the light returned from the membrane to recover the displacement spectrum of the membrane's surface.
The membrane-electrode distance is determined using the white light interferometry functionality of the same device. 
Interference fringes are collected from the partly reflective membrane layer and the electrode layer underneath, which is then used to determine their relative distance. 

The home made Michelson interferometer consists of two optical paths, namely the beam sent on the membrane and on a piezo-controlled mirror in the reference arm. 
The beams are re-combined and the relative phase measured with a high-bandwidth (0-75MHz) InGaAs balanced-homodyne receiver.
Shot noise limited measurement with an overall quantum efficiency of $25\%$ (losses, visibility, detector efficiency) is achieved.
The slow signals from the two DC monitoring outputs of the detector are used to generate the differential error signal, which is then fed to the piezo for locking the interferometer to the midpoint (maximum slope) of the interference pattern.
The rf output of the detector is high-pass filtered and fed to a spectrum analyzer in order to record the vibrations of the membrane.
 Absolute calibration of the mechanical amplitude is carried out via a calibration peak generated by driving the piezo with a known voltage at a frequency close to the mechanical peak.
The generated signal power is then converted to displacement by referring to the visibility equation and full-fringe voltage measurement determined by a slow piezo scan.

\vspace{0.5cm}
{\bf Acknowledgments}
\vspace{0.5cm}

This work was supported by the DARPA project QUASAR, the European Union Seventh Framework Program through SIQS (grant no. 600645), the ERC grants QIOS (grant no. 306576) and INTERFACE (grant no. 291038).
 We would like to thank J{\"o}rg Helge M{\"u}ller for valuable discussions, Andy Barg and Andreas N{\ae}sby for assistance with the interferometer, and Louise J{\o}rgensen for  cleanroom support.

\bibliographystyle{naturemag}

\clearpage


\setcounter{page}{0}%
\setcounter{figure}{0}%
\setcounter{equation}{0}%
\setcounter{section}{0}
\renewcommand \theequation {S\arabic{equation}}%
\renewcommand \thefigure {S\arabic{figure}}
\renewcommand \thesection {S \arabic{section}}
\renewcommand \thepage {S\arabic{page}}

\vspace{.2in}

\bibliographystyle{plain}
\begin{widetext}

\thispagestyle{empty}

\begin{center}
\large{
\textbf{ Optical detection of radio waves through a nanomechanical transducer: Supplementary Information }}
\end{center}
\vspace{.2in}

\tableofcontents

\newpage

\section{Experimental details }

\subsection{The $LC$ resonance circuit }

\subsubsection{Design}

The detailed circuit diagram we use for our experiments is sketched in Fig.S1. All the electronic components for controlling the coupling are placed on a PCB (Printed Circuit Board) that is shielded with a metal box. The output of the PCB goes to the electrical feedthroughs of the vacuum chamber in which the membrane-capacitor system is placed.  
\begin{figure*}[ht!]
    \centerline{\includegraphics[width=.7\linewidth]{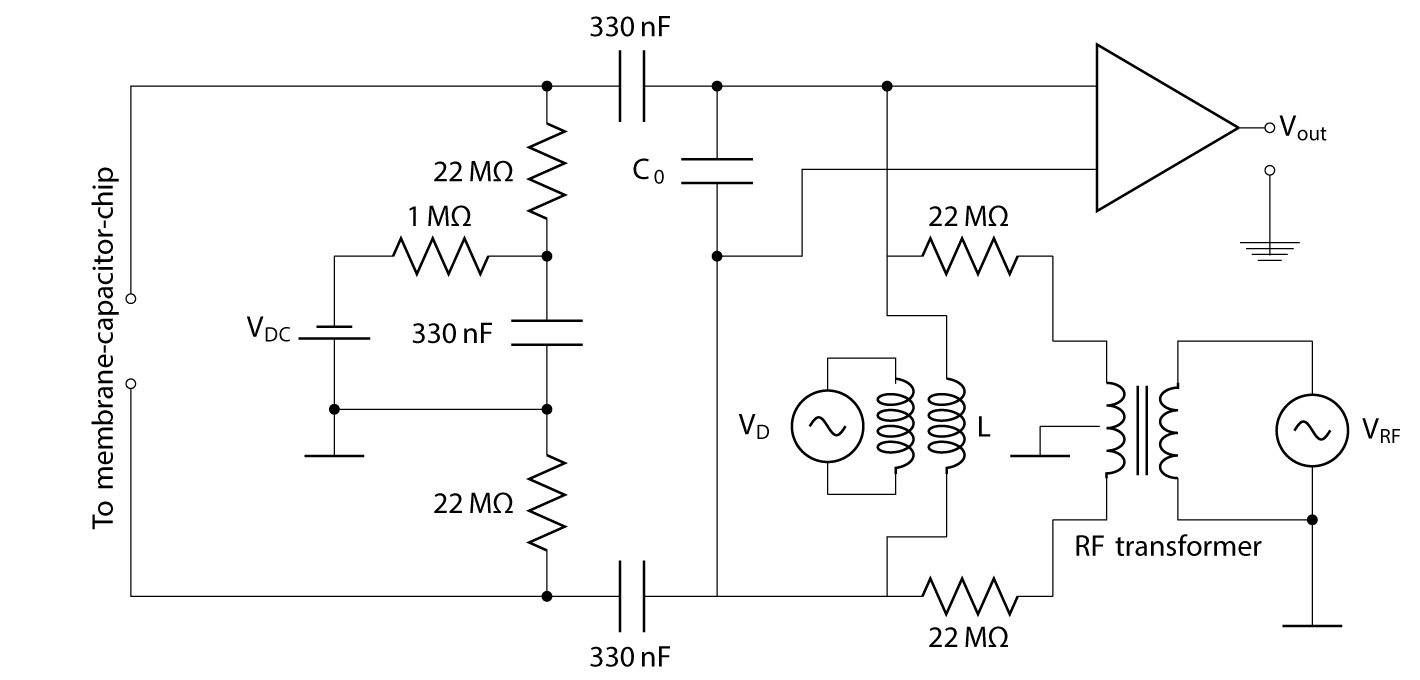}}
    \caption[diagram] {\label{fig:circuit} \textbf{Detailed circuit diagram.} Both the DC and rf voltages are applied in a symmetric way in the circuit for common-mode noise cancellation. The DC bias voltage is applied through a $1M\Omega$ resistor and a 330nF capacitor that form a low-pass filter. The two $22M\Omega$ resistors prevent loading the Q of the circuit (both for DC and rf voltage sources). The two 330nF capacitors are placed to avoid short-circuiting the DC bias through the inductor. $C_{0}$ represents the tuning capacitor to match the frequency of the LC and the mechanical resonator. For most of our measurements, the LC is driven via the drive coil in the vicinity of the ferrite inductor and the voltage on the capacitor is probed by the FET (Field Effect Transistor) operational amplifier. An additional rf port with a 1:1 transformer enables us to drive the system  when the ferrite inductor is disconnected (coupling off). For the specific measurement in Fig.1 (main text), the circuit is driven via this rf port on the PCB and probed by an additional pick-up coil.  In the real circuit, the amplifier part consists of two identical operational amplifiers taking the differential input (gain=100) followed by a third operational amplifier (gain=10), in total yielding a gain of 1000.}
\end{figure*}  

\subsubsection{Losses}
The decay rate of the LC circuit is a crucial parameter for reaching the strong coupling regime since the coupling rate should exceed this rate. In our setup, we use a high-Q ferrite inductor which is made of Ferrite 61 material (including a mixture of Zinc and Nickel) that is known for minimized core and eddy current losses. The inductor is wound with Litz wires of optimal wire diameter for low loss in the frequency range of interest ($\approx700$ kHz). Even though the Q of our inductor reaches 500 around this frequency range, our combined circuit exhibits lower Q due to several elements. We have observed that connection of the PCB (dielectric losses) and nearby magnetic elements (optical table, shielding box etc) contribute to the reduction of the Q-factor. A significant reduction occurs when the whole LC circuit is connected to the membrane-capacitor chip inside the chamber. This comes mainly from the contact resistance ($\approx50$ $\Omega$) of the gold electrode connection lines deposited on the capacitor chip. 
\begin{figure}[h!]
    \centerline{\includegraphics[width=0.5 \linewidth]{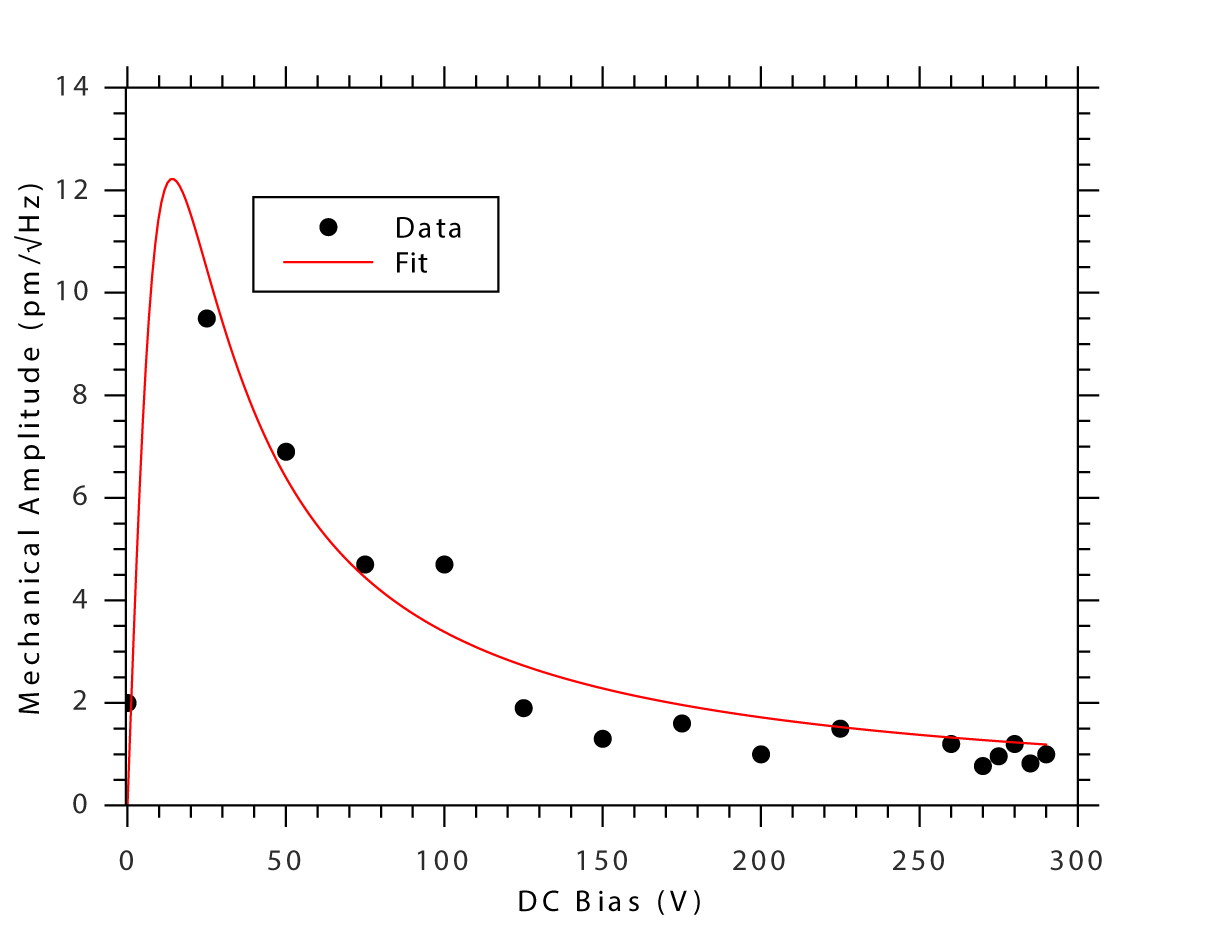}}
    \caption[noisescale] {\label{fig:excessnoise} \textbf{Optical detection of the ambient rf noise. Scaling with DC bias voltage.} The non-driven mechanical vibration amplitudes of the membrane are optically detected via FFT of the Doppler Vibrometer. The points up to 150V cover the MIT regime where the amplitude corresponds to a single resonance peak. Starting from 150V (the onset of strong coupling), we use the amplitude in the plateu between the two split-peaks. Amplitude values are extracted from Lorentzian fits to the data.}
\end{figure}   

\subsubsection{Ambient rf radiation}

In the MHz range, antennas pick up radiation that permeate the earth's atmosphere, caused mainly  by lightnings, but also man-made, and at higher frequencies, galactic sources \cite{SIKraus1966}.
Litz wires shielded with aluminium foil are used to reduce pick-up in the cables ($\sim 50 \,\mathrm{cm}$) connecting the box containing the printed circuit board control unit with the inductor.
The inductor/antenna is placed in a metal box to reduce the amplitude of the large atmospheric pickup in order to better demonstrate the sensitivity of the transducer. 
Still, the radiation picked up generates voltages largely exceeding the Johnson noise in the circuit.
We have measured the scaling of the mechanically transduced pickup signals as a function of bias voltage (Fig. \ref{fig:excessnoise}).
The noise amplitude shows scaling which would be expected for a signal coupled through the inductor port according to our model. 
The nonlinear function used for the fit is described by $f=\frac{\alpha V_\mathrm{dc}^2 V_s}{(\alpha^2 V^2+Lm\Gamma_{LC}\Gamma_m\Omega^2)}$ where $\alpha$ is a proportionality constant for $G=\alpha V_\mathrm{dc}$. The function is derived from eq.\ (\ref{e:displacement}) for the case of resonant driving ($\Omega_m=\Omega_{LC}$) with the assumption that  white noise ($V_s$) is the dominant source throughout the relevant frequency range.

\subsection{Membrane-capacitor system}

\subsubsection{Geometry and distance}
The capacitive transducer in our setup consists of a 4-segment electrode and a metal coated membrane. Microscope images of both the chip and the membrane are shown in Fig.S3. The chip has an area roughly four times larger than that of the membrane. The two pairs of segments are connected via extra electrodes such that the diagonal parts form (+) and (-) electrodes. 
We note that the overlap between the membrane area and the 4-segment electrodes after the assembly is not perfectly symmetric. 
\begin{figure}[tbh]
    \centerline{\includegraphics[width=0.7 \linewidth]{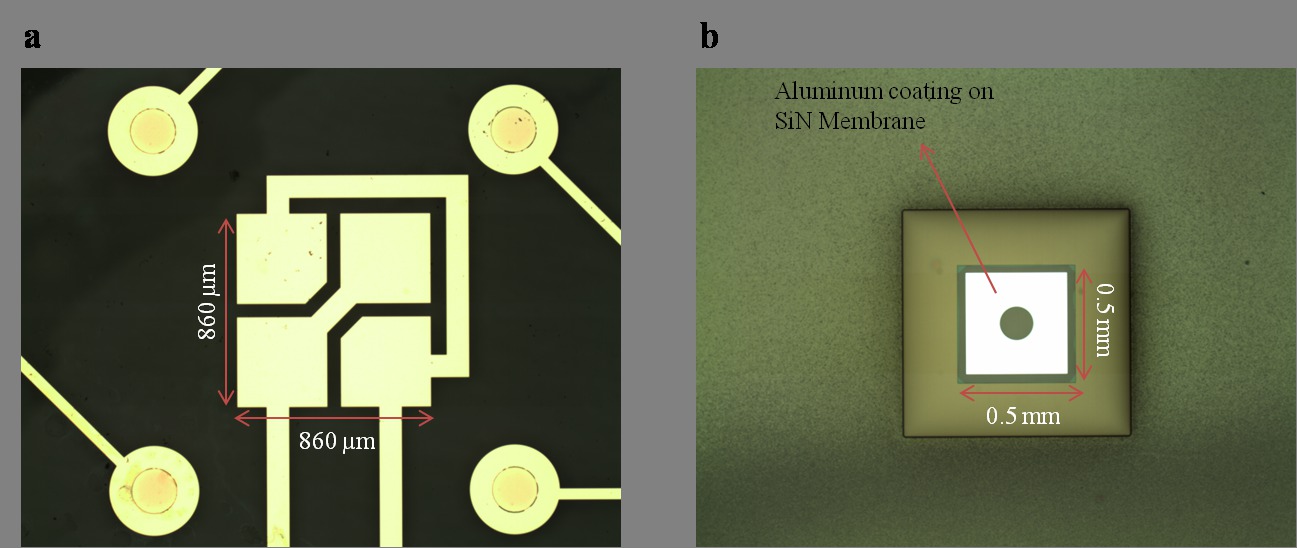}}
    \caption[membranereal] {\label{fig:microscope} \textbf{ Microscope images of the capacitor chip and membrane.} a. The capacitor chip has 4 segments. The lateral dimensions are 860$\mu$m*860$\mu$m.  The gap between the metal electrodes is 60$\mu$m.   b. SiN Membrane covers an area of 0.5 mm*0.5 mm with an Aluminum coating avoiding the edges. Rest of the large area is the frame made of Silicon.  }
\end{figure}

The initial distance between the membrane and the capacitor chip is paramount in order to achieve high coupling rates at small DC bias voltages. 
For our setup, the distance is set by the pillar height (600nm and $1\mu m$) deposited on the glass chip. 
However, most membrane-chip samples showed distances of a few $\mu m$.
This is attributed to the unevenness of the membrane and chip surface, as well as residues (silicon, resist, dust) acquired during the assembly process.
To overcome this problem, we used an extra etching process for a batch of our membranes to remove a 50 $\mu m$ thick layer of a large area within the membrane frame surface. 
This technique helped us reduce the distance closer to the pre-determined value.
As shown in Fig.\ \ref{fig:whitelight}, the distance was experimentally determined using white light interferometry.
\begin{figure}[tbh]
    \centerline{\includegraphics[width=0.5 \linewidth]{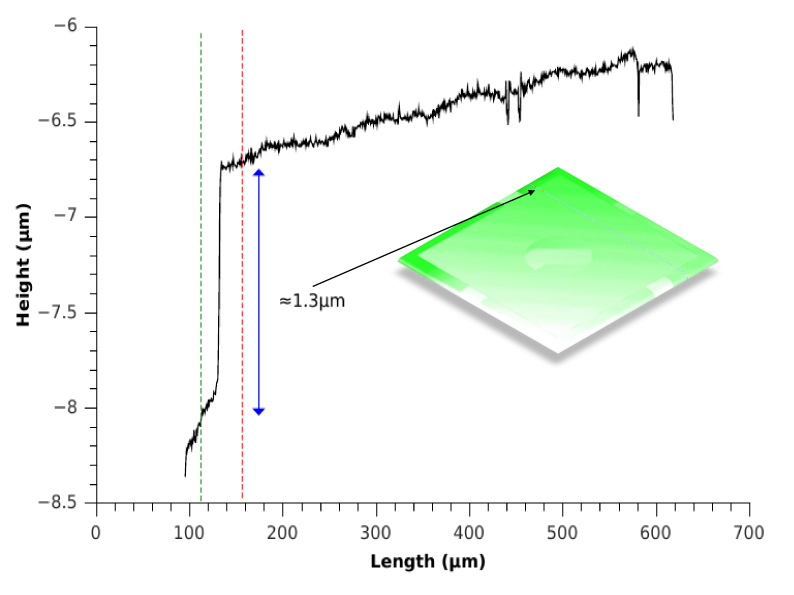}}
    \caption[distance] {\label{fig:whitelight} \textbf{White Light Interferometry for initial distance measurement.} Surface scan of a membrane-chip sample. The plot shows the surface height variation along the corresponding line-cut. Green dashed line refers to the interference pattern coming from the reflected beam from the gold electrode (beneath the semi-transparent SiN layer) whereas the red dashed line refers to the value for the Al-coated part. The difference between the two layers is a measure of the distance between the membrane and the chip-electrode around that region.}
\end{figure}   

\subsubsection{Membrane damping}

After assembly of the membrane-electrode system, we typically measured the linewidth of the thermally driven mechanical mode using the optical readout.
Typically values of $1-2\,\mathrm{Hz}$ were observed.
For some membranes, we used ringdown measurements for a more accurate determination of the mechanical quality factor.
To that end, the membrane was excited through the capacitor using a small applied voltage.
The membrane motion was again monitored optically, and its decay recorded after the excitation was switched off.
Figure \ref{fig:ringdown} shows an example of such a measurement, with a fitted (amplitude) ringdown time of $0{.}21\,\mathrm{s}$, corresponding to a linewidth of $1{.}5\,\mathrm{Hz}$ and $Q_\mathrm{m}=495000$.

When arranging the membrane very close to the electrode surface ($d\lesssim 1\,\mathrm{\mu m}$), we typically found a reduction of the membrane quality factor, with mechanical linewidths often on the order of $20\,\mathrm{Hz}$.
The precise origin of this increased damping sill has to be better understood, but could possibly arise from surface charge-membrane interaction or, for very small distances, residues getting in contact with the membrane.
The increased damping limits the cooperativity in measurements with the highest coupling, such as the one shown in Fig.\ 1 of the main manuscript.

\begin{figure}[tbh]
   \centerline {\includegraphics[width=0.4 \linewidth]{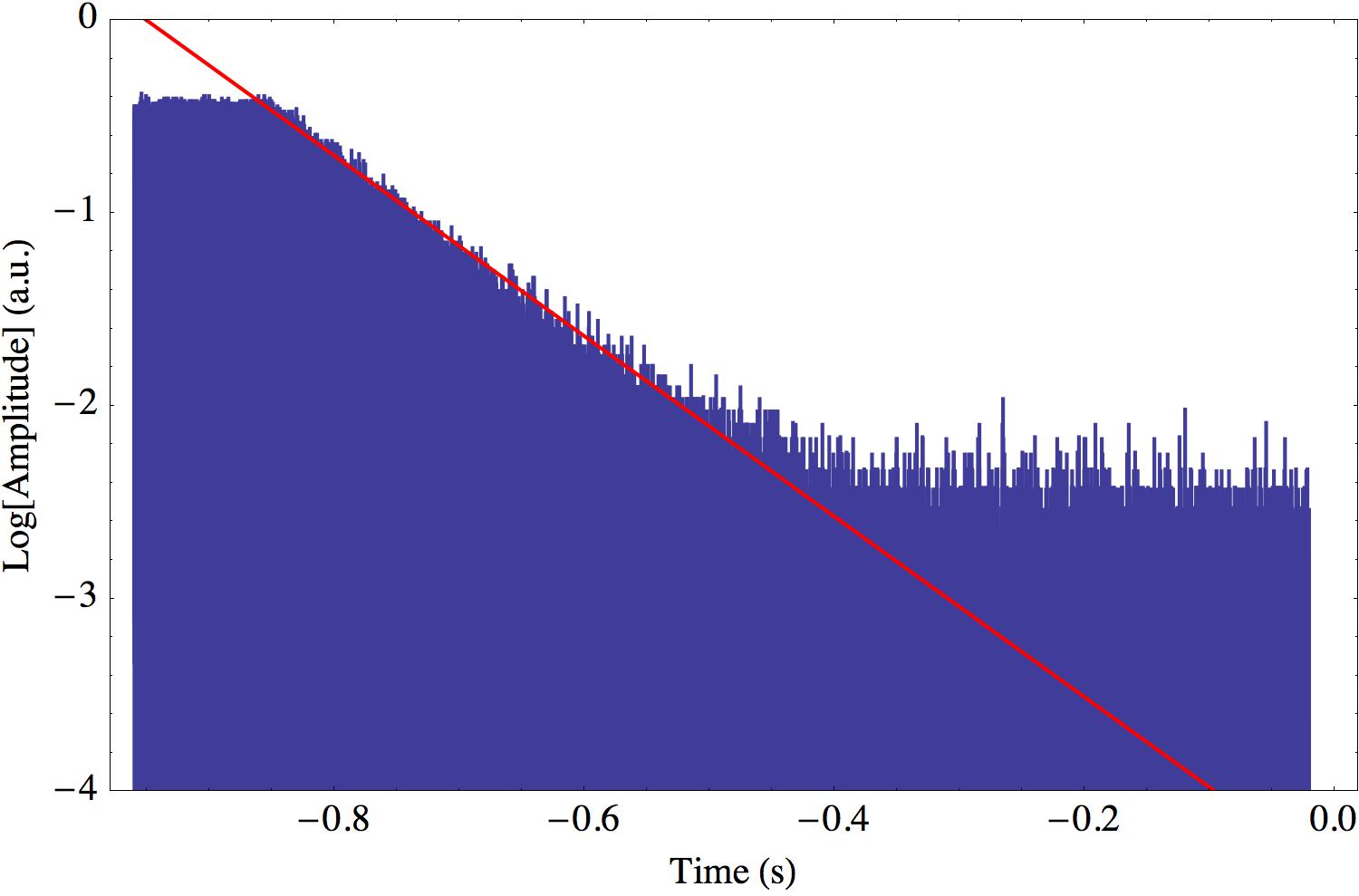}}
   \caption  {
     \label{fig:ringdown} 
     \textbf{Ringdown measurement of an Al-coated membrane.} The membrane is excited electrically and its displacement read out using optical interferometry. At $t\approx-0{.}9\,\mathrm{s}$, the excitation is stopped and the amplitude of the membrane decays exponentially. 
   Blue curve is the natural logarithm of the absolute magnitude of the voltage signal measured in the optical interferometer. Reduction by one unit on this scale indicates the  ringdown time of $0{.}21\,\mathrm{s}$, corresponding to a linewidth of $1{.}5\,\mathrm{Hz}$ and $Q_\mathrm{m}=495000$. Red line is a fit to the envelope.
     }
\end{figure} 

\subsubsection{Cooperativity limit due to instability}
In this section, we explore the limits of our device based on the maximum achievable cooperativity. Cooperativity ($\mathcal{C}_\mathrm{em}  \propto V^2$) is in principle limited by the maximum applicable DC voltage that yields a static displacement of the membrane towards the electrode direction. This voltage, known as the pull-in voltage for capacitive systems\cite{SIChowdhury2005}, can be calculated by using the scaling of the capacitive force ($F_C\propto 1/d^2$) and the equation $\frac{dF}{dx}=\frac{2kx}{d-x}-k$ at the equilibrium point, where $k$ and $d$ refer to the intrinsic spring constant and membrane-electrode distance, respectively. This expression for the electrostatically modified spring constant goes to zero if the membrane is displaced by $x=d/3$ due to the applied DC voltage, causing instability and eventual collapse of the membrane on the capacitor chip. Considering this limit, we implement the following scaling argument for our experimental data in Fig.4. Assuming a homogenous force and neglecting the deflection geometry of the membrane, frequency shift and static displacement can be related by $\frac{\Delta f}{f}\approx \frac{\Delta x}{x}$ which sets the maximum allowable frequency shift as $\Delta f_{max}=260$ kHz where $f=780$ kHz is the bare membrane frequency. Given the empirically found frequency shift of 88 kHz with 21V DC bias at a distance of $1\mu $m and the frequency shift scaling $\Delta f \propto V^2$ , the applied voltage can be increased by a factor of $\sqrt{260/88}$  which leads to a 3 times improved cooperativity of $\mathcal {C}_\mathrm{em} \approx 20000$. Consequently, this would correspond to a sensitivity limit of 2.9 pV$/\sqrt{Hz}$.

\section{Theory}

\subsection{Electro-mechanical coupling}\label{sub:Theory_EMcoupling}

\subsubsection{Hamiltonian}

The Hamiltonian \cite{SITaylor2011}
\begin{equation}
   H=\frac{\phi^2}{2 L}+\frac{p^2}{2 m}+\frac{m \Om^2 x^2}{2}+\frac{q^2}{2 C(x)}-q V
\end{equation}
describes the electromechanical system in terms of the flux $\phi$ in the inductor, the charge $q$ on the capacitors, as well as the displacement $x$ and momentum $p$ of the considered fundamental membrane mode. 
In this  scalar description the coupling arises due to the displacement dependence of the capacitance $C(x)$.

\subsubsection{Equations of motion}

The Hamiltonian directly leads to the Langevin equations 
\begin{align}
  \dot x&=\frac{p}{m}\\
  \dot p&=-m \Om ^2 x - \frac{q^2}{2} \frac{\partial}{\partial x} \left(\frac{1}{C(x)}\right)- \Gm p - F\\ 
  \dot q &= \frac{\phi}{L}\\
  \dot \phi &= 
  			-\frac{q}{C(x)}-\GLC \phi+V
\end{align}
in which the dissipation rate $\Gamma_\mathrm{m}$ of the membrane, and $\Gamma_{LC}=R/L$ of the resonance circuit have been included, as well as driving terms $F$ (a force on the membrane) and $V$ (a voltage induced in the circuit).

Assuming that the only acting force is the thermal Langevin force $\dhFth$, and writing the applied voltage as a large d.c.\ offset and a small fluctuating input 
\begin{align}
   V(t)&=V_\mathrm{dc}+ \delta V(t)
\end{align}
we can linearize the Langevin equations around an equilibrium $(\bar x, \bar p, \bar q, \bar \phi)$ characterised by
\begin{align}
  m \Om ^2 \bx 
  	&= - \frac{\bq^2}{2}  \left. \frac{\partial}{\partial x} \left(\frac{1}{C(x)}\right)\right|_{x=\bx} 
  	=  \frac{\bq^2}{2} \frac{C'(\bx)}{C(\bx)^2}\\ 
  \bq &=  V_\mathrm{dc} C(\bx)\\
  \bar p &= \bar \phi = 0
\end{align}
and, to first order in the small fluctuations (provided the system is stable),
\begin{align}
  \dot \dhx(t)&=\frac{\dhp(t)}{m}\\
  \dot \dhp(t)&=
  	-m \Om ^2 \dhx(t) 
  	-\underbrace{\frac{\bq^2}{2} \left. \frac{\partial^2}{\partial x^2} \left(\frac{1}{C(x)}\right)\right|_{x=\bx}}_{2 m \Om \,\Delta\Om} \dhx(t)
  	- \Gm p - \underbrace{\bq \left. \frac{\partial}{\partial x} \left(\frac{1}{C(x)}\right)\right|_{x=\bx}}_{G}  \dhq(t)- \dhFth(t)\\ 
  \dot \dhq(t) &= \frac{\dhphi(t)}{L}\\
  \dot \dhphi(t) &= 
  	-\frac{\dhq(t)}{C(\bx)}
	- \underbrace{\bq \left. \frac{\partial}{\partial x} \left(\frac{1}{C(x)}\right)\right|_{x=\bx}}_{G} \dhx(t)
	-\GLC \dhphi(t)+ \delta V(t)
\end{align}
Here, we have introduced the coupling parameter
\begin{equation}\label{e:G_def}
  G	=\bq \left. \frac{\partial}{\partial x} \left(\frac{1}{C(x)}\right)\right|_{x=\bx}
  	=-\bq \frac{C'(\bx)}{C(\bx)^2}
\end{equation}
as well as the frequency shift
\begin{equation}
  \Delta\Om	=\frac{\bq^2}{4 m \Om} \left. \frac{\partial^2}{\partial x^2} \left(\frac{1}{C(x)}\right)\right|_{x=\bx}.
\end{equation}

Absorbing the frequency shift into a re-defined $\Om$ and transforming to the Fourier domain yields
\begin{align}
- i \Og \, \dhx(\Og) &=\dhp(\Og) /m 									\label{e:eomlfx}\\
- i \Og \, \dhp(\Og) & = - m \Om^2 \,\dhx(\Og) - \Gm \,\dhp(\Og) - G \, \dhq(\Og) - \dhFth(\Og) 	\label{e:eomlfp}\\
- i \Og \, \dhq(\Og)  & = \dhphi(\Og) / L 								\label{e:eomlfq}\\
- i \Og \, \dhphi(\Og) & = -\dhq(\Og)/C- \GLC \, \dhphi(\Og) - G \, \dhx(\Og) - \dhV(\Og) 	\label{e:eomlfphi}.
\end{align}
These algebraic equations can be used to calculate the response of the system to excitations through a force or voltage drive.
For notational convenience, we often use the bare susceptibilities
\begin{align}
  \chim(\Og)&=\frac{1}{m\left(\Om^2-\Og^2-i \Og \Gm\right) }\\
  \chiLC(\Og)&=\frac{1}{L\left(\OLC^2-\Og^2-i \Og \GLC\right)}.
\end{align}
of the mechanical and $LC$ resonator, respectively.

\subsubsection{Coupling rate and cooperativity}

The system (\ref{e:eomlfx})-(\ref{e:eomlfphi}) represents two oscillators coupled with a coupling energy 
\begin{equation}
 \hbar g_\mathrm{em}=G x_\mathrm{zpf} q_\mathrm{zpf}=G \sqrt{\frac{\hbar}{2 m \Om }} \sqrt{\frac{\hbar}{2 L \OLC}},
\end{equation}
where $x_\mathrm{zpf}$ and $q_\mathrm{zpf}$ are the zero-point fluctuations of the membrane displacement and capacitor charge, respectively.
Comparison of the corresponding coupling rate $g_\mathrm{em}$ with the dissipation rate yields the electromechanical cooperativity \cite{SIGroblacher2009a,SIVerhagen2011}
\begin{equation}
\mathcal{C}_\mathrm{em}= \frac{4 g_\mathrm{em}^2}{\Gm \GLC}.
\end{equation}


\subsection{Probing the system}

\subsubsection{Voltage across the capacitor}

The voltage across the capacitor(s), measured at port ``1'' of the circuit, is given by
\begin{align}
  V_\mathrm{c}&=\frac{q}{C(x)}
  \intertext{and further}
  \bar V_\mathrm{c}+\delta V_\mathrm{c}&=(\bq+\dhq)\left(\frac{1}{C(\bx)}+ \left. \frac{\partial}{\partial x} \frac{1}{C(x)}\right|_{x=\bx} \dhx \right), \nonumber\\
  \delta V_\mathrm{c}		  						&= \frac{\dhq}{C(\bx)}+ G \dhx 
 \label{e:dvc}
\end{align}
after again applying linearisation.
Together with 	(\ref{e:eomlfx})-(\ref{e:eomlfphi}) this yields 
 \begin{align}
  \delta V_\mathrm{c}(\Og) 	  	&= \left(\frac{1}{C(\bx)}- \chim(\Og)G^2   \right)\dhq(\Og) = \nonumber \\
  								&= \left(\frac{1}{C(\bx)}- \chim(\Og)G^2   \right)\underbrace{\left( \chiLC(\Og)^{-1} - G^2 \chim(\Og)   \right)^{-1}}_{\chi_\mathrm{LC}^\mathrm{eff}(\Og) }(- \dhVs (\Og)),
\label{e:MITcapacitor}
\end{align}
if we assume the voltage is most significantly affected by a voltage $\dhVs$ induced in the circuit, which is the case when the circuit is driven through a sufficient strong voltage pickup (port ``2'').
Figure \ref{f:MITcapacitor} shows typical expected response curves, showing the transition from mechanically induced transparency to  normal mode splitting.

\begin{figure}[tbh]
  \centering
  \includegraphics[width=.5\linewidth]{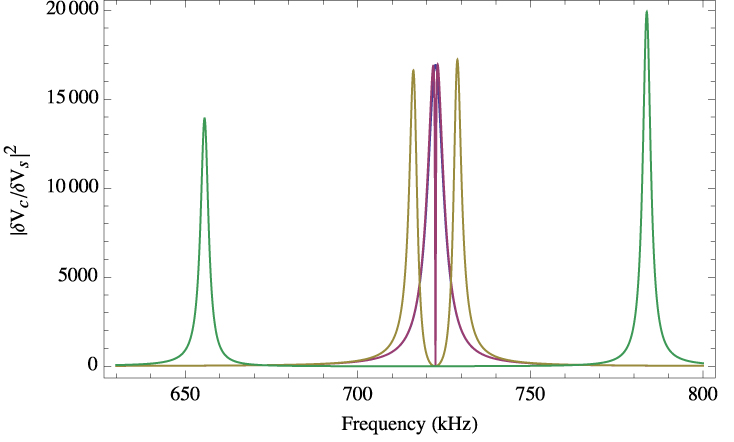}
   \caption{
   Mechanically induced transparency, and modal splitting for strong coupling, as observed in the voltage across the capacitor(s).
   Numerical parameters are typical for our experiments, $\Om=\OLC=2 \pi\cdot 722 \,\mathrm{kHz}$, $\Gm=2 \pi\cdot  1.5 \,\mathrm{Hz}$, $\GLC=2 \pi\cdot  722 \,\mathrm{kHz}/130= 2 \pi\cdot  5.5 \,\mathrm{kHz}$, $m=  28 \,\mathrm{ng}$, $ C=  72 \,\mathrm{pF} $, $L=  675 \,\mathrm{\mu H}$, and   $G=\{0, 5, 50, 500\} \,\mathrm{kV/m}$ for $\{$blue, violet, yellow, green$\}$ curves, in the right panel, the transmission on resonance
   $\Og=\Omega_\mathrm{R}$ is shown as a function of $G$.}
   \label{f:MITcapacitor}
\end{figure}

\subsubsection{Optical measurement of the membrane}

The optical field reflected from the membrane experiences a phase shift of 
\begin{align}
  \delta \varphi_\mathrm{mem}&=2 k \dhx 
\end{align}
 where $k=2 \pi/\lambda$ is the readout light's wavenumber.
From  (\ref{e:eomlfx})-(\ref{e:eomlfphi})  we have
\begin{equation}
  \underbrace{
  \left(\chim^{-1}(\Og)-G^2  \chiLC(\Og)\right)
  }_{\chi_\text{m, eff}^{-1}(\Og)} \dhx(\Og)=-\dhFth(\Og)+G \chiLC(\Og) \dhV(\Og)
  \label{e:displacement}
\end{equation}
for the mechanical displacement, so that the total measured optical phase shift is
\begin{align}
  \delta \varphi(\Og)&=2 k \chi_\text{m}^\mathrm{eff}(\Og) \left[ -\dhFth(\Og)+G \chiLC(\Og) \dhV(\Og) \right] + \delta \varphi_\mathrm{im}(\Og),
    \label{e:phaseshift}
\end{align}
where we have taken the  imprecision $\delta \varphi_\mathrm{im}$ of the measurement into account, but neglected optical backaction \cite{SIPurdy2013}, which would only become relevant for much greater optical powers than the ones used here ($\lesssim 10 \,\mathrm{mW}$).

\subsection{Sensitivity}
\label{Sensitivity}
From eq.\ (\ref{e:phaseshift}), the spectral density of phase fluctuations is given by
\begin{equation}
 S_{\varphi\varphi}^\mathrm{tot}(\Omega)=  (2k)^2\left| \chi_\mathrm{m}^\mathrm{eff}(\Omega) \right |^2 \left(
  				\left| G \chi_{LC} (\Omega) \right |^2 S_{VV}(\Omega)+S_{FF}^\mathrm{th} (\Omega)\right)+S_{\varphi\varphi}^\mathrm{im}(\Omega).
\end{equation}
for uncorrelated inputs $\delta V$, $\delta F_\mathrm{th}$ and $\delta \varphi_\mathrm{im}$.
A voltage signal can therefore be detected as soon as its spectral density exceeds both the optical readout noise
\begin{equation}
 S_{VV}^\mathrm{opt}(\Omega)=\frac{S_{\varphi\varphi}^\mathrm{im}(\Omega)}{ \left| 2 k  \chi_\mathrm{m}^\mathrm{eff}(\Omega)  G \chi_{LC} (\Omega) \right |^2},
\end{equation}
as well as the noise due to thermal fluctuations of the membrane position
\begin{equation}
 S_{VV}^\mathrm{mem}(\Omega)=\frac{S_{FF}^\mathrm{th} (\Omega)}{ \left|  G \chi_{LC} (\Omega) \right |^2}.
\end{equation}
The condition $ S_{VV}^\mathrm{opt}(\Omega)<< S_{VV}^\mathrm{mem}(\Omega)$ can be achieved by increasing the optical power and thus reducing the shot noise limit of $S_{\varphi\varphi}^\mathrm{im}(\Omega)$. Underthese conditions, with the spectral density of the Langevin force $S_{FF}^\mathrm{th}(\Og)=2 m \Gm k_\mathrm{B} T_\mathrm{m}$, the fundamental limit for the sensitivity can be written as 
\begin{equation}
 S_{VV}^\mathrm{mem}(\Omega)=\left|  \frac{\chi_{LC} (\OLC)}{\chi_{LC} (\Omega)} \right |^2 \cdot  2 k_\mathrm{B} \frac{T_\mathrm{m}}{\mathcal{C}_\mathrm{em}} R.
\end{equation}

\subsection{Modulation and conversion efficiency}

In the context of efficient light modulation by an RF field, the coupling strength can be quantified in terms of the voltage $V_\pi$, which, if applied to the LC circuit, shifts the reflected light's phase by $\pi$.
Focusing on the resonant response, we find from eq.\ (\ref{e:phaseshift})
\begin{align}
    \delta \varphi(\Omega_\mathrm{R})&=2 k \sqrt{\frac{C}{m\GLC\Gm}\frac{\mathcal{C}_\mathrm{em} }{(\mathcal{C}_\mathrm{em} +1)^2}} \dhV(\Omega_{R}),
  \end{align}
which is maximum at $\mathcal{C}_\mathrm{em} =1$, yielding 
\begin{align}
   V_\pi&=\frac{\lambda}{2}\sqrt{\frac{m \GLC \Gm}{C}}=\frac{\lambda}{2}\sqrt{m L \GLC \Gm} \Omega_\mathrm{R}
  \end{align}
 in this case.
 
 It is straightforward to estimate the conversion efficiency of radio-frequency photons to optical sideband photons of the light reflected from the membrane,
\begin{align}
   \eta_\mathrm{eo}	=\frac{P_\mathrm{sbs}/\hbar \omega_\mathrm{opt}}{P_\mathrm{RF} /\hbar \Omega_\mathrm{R}}
   					=\frac{\Phi_\mathrm{car} \cdot2 \left(\frac{\delta\varphi }{2}\right)^2}{|\delta V_\mathrm{s} \delta I_{s}|/\hbar \Omega_\mathrm{R}}
					=\frac{\Phi_\mathrm{car} \cdot2 \left(\frac{\pi }{2}\frac{\delta V}{V_\pi}\right)^2}{|\delta V_\mathrm{s}  \, \Omega\, q|/\hbar \Omega_\mathrm{R}}
					=\frac{4 C_\mathrm{em}}{1+C_\mathrm{em}}(k x_0)^2 \frac{ \Phi_\mathrm{car}}{\Gm}.
\end{align}
Note that as opposed to $V_\pi$ where $C_\mathrm{em}=1$ was found to be optimum, high conversion efficiencies can be achieved for large cooperativities as the impedance of the circuit is increased and thus less power is provided by the source for a given source voltage $V_\mathrm{s}$.

For sufficiently high cooperativity, ${4 C_\mathrm{em}}/(1+C_\mathrm{em})\rightarrow 4$, and with $m=30\,\mathrm{ng}$, $\Omega_\mathrm{m}/2\pi=690\,\mathrm{kHz}$, $\Gamma_\mathrm{m}/\pi=2\,\mathrm{Hz}$, $\Phi_\mathrm{car} h c/\lambda=20\,\mathrm{mW}$, we find $ \eta_\mathrm{eo}=48\%$.
 
\section{Data Analysis}

\subsection{Fit models}

\subsubsection{Membrane thermal noise spectrum}

The bare membrane mechanical characteristics can be extracted from a measurement of its displacement when thermally driven only, in absence of electromechanical coupling  ($V_\mathrm{dc}=0$).
To this end, we use a Doppler vibrometer  which provides a calibrated displacement spectrum.
A simple Lorentzian model
\begin{equation}
 S_{xx}(\Og)=|\chim(\Og)|^2 S_{FF}(\Og)=\frac{1}{m^2}\frac{1}{(\Om^2-\Og^2)^2+\Gm^2 \Og^2}2 m \Gm \kB T_\mathrm{m}
\end{equation}
is fit to this spectrum and used to extract the mechanical damping $\Gm$ and the effective mass $m$ (assuming $T_\mathrm{m}=300\,\mathrm{K}$).
Note that this mass is an \emph{effective optical mass}, that is, it may vary slightly depending on the probing point of the optical beam.
The variations we observe are on the order of $20\%$.

\subsubsection{Spectrum of voltage fluctuations across the capacitor}

The response of the electromechanical systems to a series voltage picked up in the inductor (port ``2''), and measured within the circuit, at the capacitor (port ``1''), is described by eq.\ (\ref{e:MITcapacitor}).
Here, we have assumed that the signal is dominated by the induced voltage. 
The corresponding model
\begin{align}
  \delta V_\mathrm{c}(\Og)  &= -\left(\frac{1}{C(\bx)}- \chim(\Og)G^2   \right){\chi_\mathrm{LC}^\mathrm{eff}(\Og) } \dhVs (\Og)
  \label{e:fitmodelV}
\end{align}
is fit to the response functions measured in this fashion, and used to extract the parameters $\Om$, $\OLC$, $\GLC$, $G$ and $\dhVs$ (the amplitude of the induced driving voltage).
The required total capacitance has been determined beforehand in independent measurements, and is typically at a level of $C(\bar x)=80\,\mathrm{pF}$.

Note  that in this approach, the electromechanical coupling parameter $G$ depends indirectly on the (optically determined) effective mass $m$, as the actual electromechanical coupling physics (and thus the shape of the spectra, e.\ g.\ splitting) is determined by the coupling energy $\hbar g\propto G/\sqrt{m}$, which is independent of the optical probing point.

\subsubsection{Spectrum of the optical phase shift}

The response of the optical phase shift to a voltage pickup in port ``2'' is readily described by eq.\ (\ref{e:phaseshift}), in the limiting case where the driving voltage dominates the signal,
\begin{align}
  \delta \varphi(\Og)&=2 k \chi_\text{m}^\mathrm{eff}(\Og) G \chiLC(\Og) \dhVs(\Og).
\end{align}
By performing a calibration of the spectra into mechanical displacement units, a simplified fit model of
\begin{align}
  \dhx(\Og)&=\chi_\text{m}^\mathrm{eff}(\Og) G \chiLC(\Og) \dhVs(\Og).
  \label{e:fitmodelX}
\end{align}
can be used, with input parameters $m$ and $C$ from independent measurements, as well as $\GLC$ as determined by the fits of the voltage measurements at port ``1'' (see above).
The  parameters $\Om$, $\OLC$,  $G$ and $ \dhVs$ are adjusted by the fit.
The results for  $\Om$, $\OLC$, and, importantly,  $G$, typically agree to the fits to the voltage spectra to within $1\%$. 

\subsection{Calibration of the coupling rate $G$}

\subsubsection{Spectral shape}

As described in the previous section, fits to the spectra of the intracircuit voltage (port ``1'') or interferometrically measured optical phase shift yield values of the coupling parameter $G$. 
Note that determination of $G$ is performed by fitting the \emph{shape} of the spectra with the amplitude as a free parameter.
The amplitudes can be used in an additional approach  to determine $G$ (see below).
The parameter $G$ can also be re-expressed into a coupling rate $g_\mathrm{em}$ using the identity
\begin{equation}
   \hbar g_\mathrm{em}=G \sqrt{\frac{\hbar}{2 m \Om}}\sqrt{\frac{\hbar}{2 L \OLC}}
\end{equation}
and thus made independent of the optical mass $m$.

For the data shown in figure 2 of the main manuscript, we find $G_\mathrm{i}=10.3\,\mathrm{kV/m}$ at $V_\mathrm{dc}=125\mathrm{V}$.

\subsubsection{Amplitude ratio}

Alternatively, the \emph{amplitudes} of  the voltage in the circuit $\delta V_\mathrm{c}$ and the mechanical displacement $\dhx$ can be compared. 
For example, at resonance, we find
\begin{equation}
 \left| \frac{\delta V_\mathrm{c}(\Omega_\mathrm{R})}{\dhx(\Omega_\mathrm{R})} \right|=\sqrt{G^2+\frac{L^2 m^2 \Gm \Omega_\mathrm{R}^6}{G^2}},
 \label{e:ampratio}
\end{equation}
encoding $G$ as a function of the experimentally measured modulation amplitudes.
In practice it turned out to be more robust to evaluate this ratio not at one particular frequency (here e.\ g.\ $\Omega_\mathrm{R}$), but throughout the entire measured frequency range.
This can be conveniently accomplished by evaluating the left-hand-side of eq.\ (\ref{e:ampratio}) using the models (\ref{e:fitmodelV}) and (\ref{e:fitmodelX}), with the parameters fitted to the experimental spectra, including---crucially---the (possibly) different drive amplitudes $\dhVs$ determined in these fits, which uniquely encode the amplitudes of the measured spectra. 

For the driven measurements with the Lock-in (in units of Vrms), we implement the following approach to calibrate the readings to mechanical amplitudes (shown in Fig.2). The membrane is first excited with a single frequency sine wave of known amplitude from the rf output of the Lock-in via a drive coil. This appears as a sharp mechanical response on the displacement spectrum as measured with the vibrometer, which is by default calibrated in meters. This value is then used to calibrate the wide membrane spectrum when we scan the rf excitation signal and record the driven response of the Vibrometer output which is fed to the Lock-in.

For the data shown in figure 2 of the main manuscript, we find $G_\mathrm{ii}=8.5\,\mathrm{kV/m}$ at $V_\mathrm{dc}=125\mathrm{V}$.

\subsubsection{Frequency shift}

A third, independent approach to estimating the coupling strength is by measuring the frequency shift when a \mbox{d.\ c.}\ voltage is applied.
Above we found %
\begin{align}
  \Delta \Om		&=\frac{\bq^2}{4 m^{*} \Om}\cdot \left. \frac{\partial^2 }{\partial x^2} \frac{1}{C(x)}\right|_{x=\bx},
  \end{align}
where we have chosen canonical coordinates such that the oscillator mass equals the physical membrane mass $m^{*}$.
To relate the frequency shift and the coupling rate, a relation of the first and second derivative of the inverse capacitance at the operating point $\bar x$ is necessary.
Without loss of generality, we can write 
\begin{equation}
  \left. \frac{\partial}{\partial x} \frac{1}{C(x)}\right|_{x=\bx}=-\left. \frac{\partial^2 }{\partial x^2} \frac{1}{C(x)}\right|_{x=\bx} \cdot D
  \end{equation}
with some characteristic length $D$.
Then we obtain 
\begin{equation}
 G=\bq \left. \frac{\partial}{\partial x} \frac{1}{C(x)}\right|_{x=\bx} 
 =-\bq D \left. \frac{\partial^2 }{\partial x^2} \frac{1}{C(x)}\right|_{x=\bx}= 
	-\frac{ 4m^{*} \Om  D }{C  V_\mathrm{dc}}\Delta \Om \label{e:freqshift}.
\end{equation}
An approximation for $D$ can be obtained by assuming
\begin{equation}
  C(x)=C_0+C_\text{m}(x)=C_0+\frac{a}{d+x},
\end{equation}
where $d$ is an equilibrium distance, e.\ g.\ between the plates of a capacitor, one of which can move. 
This model yields $D=d/2$ for a large capacitance offset $C_0\gg a/(d+x)$.
More generally, for any capacitive force for which
\begin{equation}
F(x)\propto   \frac{\partial}{\partial x} \frac{1}{C(x)} \propto d^{-2}
 \end{equation}
holds approximately, one finds $D=d/2$. For a general choice of gauge for the oscillator mass $m \neq m^{*}$ (accompanied by a suitable rescaling of $x$), the chain rule yields $D=\sqrt{m/m^{*}}d/2$; thus
\begin{equation}
 G=
	-\frac{ 2\sqrt{m m^{*}} \Om  d }{C(\bx)  V_\mathrm{dc}}\Delta \Om .
\end{equation}
This expression can be used to estimate the coupling parameter $G$ by measuring the frequency shift $\Delta \Om$, if the distance $d$, effective and physical masses $m$ and $m^{*}$, total capacitance $C(\bx)$ and applied bias voltage $V_\mathrm{dc}$ are known.

For the data shown in figure 2 of the main manuscript, we find $G_\mathrm{iii}=14.4\,\mathrm{kV/m}$  at $V_\mathrm{dc}=125\mathrm{V}$.

\subsubsection{Theoretical calculation}
The total capacitance of the membrane-electrode arrangement depends on the membrane mode shape and the electrode mask as well as their relative alignment. This capacitance can be determined analytically in a quasi-electrostatic calculation by employing certain geometrically motivated approximations \cite{SICouplingFrameworkInPrep}.

We proceed by assuming that the curvature of the membrane is sufficiently small that we, for purposes of calculating the capacitance, can take it to be locally flat. If we furthermore neglect edge effects, symmetry considerations imply that we may model the membrane-electrode capacitance \emph{locally} as that of parallel plates (assumed to be perfectly conducting). This local capacitance per area only depends on the local membrane-electrode separation along the direction normal to the plane defined by the electrodes.

Within the stated approximations, we can now construct the total capacitance of the arrangement in terms of the local capacitance contributions. Noting that the membrane-electrode separation is significantly smaller than the inter-electrode gaps of the quarter-segment geometry (see Fig.~\ref{fig:microscope}a), we can neglect the direct capacitance between electrode segments. Instead, the capacitance between the positive and negative electrodes is mediated by the membrane, so that the integrated local capacitance between membrane and positive electrode segments, $C_{+}$, acts in series with the corresponding quantity for the negative electrode segments, $C_{-}$:
\begin{equation}\label{e:C_series_model}
C = C_0 + \frac{1}{\frac{1}{C_{+}} + \frac{1}{C_{-}}},
\end{equation}
where $C_0$ is the tuning capacitance acting in parallel with the membrane modulated capacitance.

With this model of the total capacitance $C$, we may proceed to calculate the coupling strength $G$ along the lines of the derivation given in subsection~\ref{sub:Theory_EMcoupling}. The displacement configuration of the membrane $\delta x(y,z)$ relative to its static deflection (induced by the DC voltage bias) can be expanded on the set of drum modes (where $y$ and $z$ are cartesian coordinates of the membrane plane):
\begin{equation}\label{e:drum_mode_expansion}
\delta x(y,z) = \sum_{m,n} \beta_{m,n} \Phi_{m,n}(y,z).
\end{equation}
The coefficients $\beta_{m,n}$ are canonical drum mode position coordinates each of which has the free evolution of a harmonic oscillator. Their frequencies are given by the drum mode spectrum, while we have a gauge freedom of choosing their masses by means of canonical rescaling of the $\beta_{m,n}$ and their conjugate momenta. This also implies a rescaling of the modes $\Phi_{m,n}(y,z)$ in order to retain the form of Eq.~(\ref{e:drum_mode_expansion}) (note that the coordinate axis for $\delta x$ is fixed once and for all in specifying the membrane-electrode distance $d$ in absence of bias voltage). That is, focusing on a single drum mode $\Phi(y,z)$ henceforth, the choice of gauge mass $m$ for the canonical variables associated with this mode entails the requirement
\begin{equation}\label{e:Phi_normalization}
\iint\limits_{A_{\text{mem}}}\! \Phi^2(y,z) \dif y \dif z = \frac{m}{m^{*}},
\end{equation}
where $m^{*}$ is the physical mass of the membrane and $A_{\text{mem}}$ is the membrane area.

In parallel to Eq.~(\ref{e:G_def}) we write
\begin{equation}\label{e:G_beta}
G=\bar{q}\frac{\partial}{\partial\beta}\left.\left(\frac{1}{C[{\delta x(y,z)}]}\right)\right|_{\text{eq}},
\end{equation}
where 'eq' signifies that the derivative should be evaluated at the static displacement equilibrium configuration of the membrane. However, assuming the static displacement to be negligible compared to $d$, it is a good approximation to evaluate Eq.~(\ref{e:G_beta}) at the flat membrane configuration. Inserting our model for $C$, Eq.~(\ref{e:C_series_model}), we find
\begin{equation}\label{e:G_beta_result}
G = \frac{V_{\text{dc}}\epsilon_0 L^2}{C^{\text{(eq)}} d^2} \sqrt{\frac{m}{m^{*}}} \left[ \frac{\frac{O^{(1)}_{+}}{[O^{(0)}_{+}]^2} + \frac{O^{(1)}_{-}}{[O^{(0)}_{-}]^2}}{\left(\frac{1}{O^{(0)}_{+}} + \frac{1}{O^{(0)}_{-}} \right)^2}\right],
\end{equation}
where $L$ is the membrane side length and we have introduced the overlap factors
\begin{equation}
O_{i}^{(j)} \equiv \frac{\iint\limits_{A_{i}}\! \Phi^j(y,z)\xi(y,z)\dif y\, \dif z}{\left(\iint\limits_{A_{\text{mem}}}\!\Phi^2(y,z)\dif y\,\dif z\right)^{j/2}},
\end{equation}
with $i\in\{+,-\}$ and $A_i$ being the area of the membrane above electrodes of polarity $i$; $\xi(y,z)$ is a product of Heaviside functions that masks the electrode gaps as well as the hole in the membrane metalization (see Fig.~\ref{fig:microscope}).

In order to evaluate $G$, we need to calculate the overlap term appearing in brackets in Eq.~(\ref{e:G_beta_result}) for our geometry. For the fundamental mode we find the value $0.128$ provided perfectly central positioning of the membrane with respect to the quarter-segment electrodes. In a range of up to 25\% lateral misalignment the value varies between $0.119$ and $0.139$ with an (unweighted) average of $0.131$, which is the value we will use below. The physical mass can be estimated to $m^{*}=110 \text{ng}$ using the mass densities $\rho_{\text{SiN}}=3.0 \text{g}/\text{cm}^3$ and $\rho_{\text{Al}}=2.7 \text{g}/\text{cm}^3$. The membrane side length $L$ is $0.5 \text{mm}$.

For the specific experiment presented in Fig. 2 of the main text, we have effective mass $m=30 \text{ng}$ and membrane-electrode separation $d=5.5 \mu \text{m}$. As the LC mode is retuned into resonance with the mechanical mode each time the bias voltage $V_{\text{dc}}$ is modified, Eq.~(\ref{e:G_beta_result}) acquires an additional $V_{\text{dc}}$ dependence through $C^{\text{(eq)}} \approx C_0$, where $C_0$ is the tuning capacitance. However, this amounts to a correction of a few percent over the voltage range considered in the aforementioned figure; for simplicity then, we may take $C_0=76 \text{pF}$ independently of voltage. Evaluating Eq.~(\ref{e:G_beta_result}) for these values, we find $G/V_{\text{dc}} = 66 \text{m}^{-1}$. For $V_{\text{dc}}=125 \text{V}$ we find a value of $G_{\text{iv}}=8.2 \text{kV}/\text{m}$.

\end{widetext}


\begin{thebibliography}{10}
\expandafter\ifx\csname url\endcsname\relax
  \def\url#1{\texttt{#1}}\fi
\expandafter\ifx\csname urlprefix\endcsname\relax\def\urlprefix{URL }\fi
\providecommand{\bibinfo}[2]{#2}
\providecommand{\eprint}[2][]{\url{#2}}

\bibitem{Kippenberg2008}
\bibinfo{author}{Kippenberg, T.~J.} \& \bibinfo{author}{Vahala, K.~J.}
\newblock \bibinfo{title}{Cavity {O}ptomechanics: {B}ack-{A}ction at the
  {M}esoscale}.
\newblock \emph{\bibinfo{journal}{Science}} \textbf{\bibinfo{volume}{321}},
  \bibinfo{pages}{1172--1176} (\bibinfo{year}{2008}).

\bibitem{Aspelmeyer2013}
\bibinfo{author}{Aspelmeyer, M.}, \bibinfo{author}{Kippenberg, T.~J.} \&
  \bibinfo{author}{Marquardt, F.}
\newblock \bibinfo{title}{Cavity optomechanics}.
\newblock \emph{\bibinfo{journal}{arXiv:1303.0733}}  (\bibinfo{year}{2013}).

\bibitem{OConnell2010}
\bibinfo{author}{O'Connell, A.~D.} \emph{et~al.}
\newblock \bibinfo{title}{Quantum ground state and single-phonon control of a
  mechanical resonator.}
\newblock \emph{\bibinfo{journal}{Nature}} \textbf{\bibinfo{volume}{464}},
  \bibinfo{pages}{697--703} (\bibinfo{year}{2010}).

\bibitem{Teufel2011}
\bibinfo{author}{Teufel, J.~D.} \emph{et~al.}
\newblock \bibinfo{title}{Circuit cavity electromechanics in the strong
  coupling regime}.
\newblock \emph{\bibinfo{journal}{Nature}} \textbf{\bibinfo{volume}{471}},
  \bibinfo{pages}{204--208} (\bibinfo{year}{2011}).

\bibitem{Palomaki2013}
\bibinfo{author}{Palomaki, T.~A.}, \bibinfo{author}{Harlow, J.~W.},
  \bibinfo{author}{Teufel, J.~D.}, \bibinfo{author}{Simmonds, R.~W.} \&
  \bibinfo{author}{Lehnert, K.~W.}
\newblock \bibinfo{title}{Coherent state transfer between itinerant microwave
  fields and a mechanical oscillator}.
\newblock \emph{\bibinfo{journal}{Nature}} \textbf{\bibinfo{volume}{495}},
  \bibinfo{pages}{210--214} (\bibinfo{year}{2013}).

\bibitem{Groblacher2009a}
\bibinfo{author}{Gr{\"o}blacher, S.}, \bibinfo{author}{Hammerer, K.},
  \bibinfo{author}{Vanner, M.~R.} \& \bibinfo{author}{Aspelmeyer, M.}
\newblock \bibinfo{title}{Observation of strong coupling between a
  micromechanical resonator and an optical cavity field}.
\newblock \emph{\bibinfo{journal}{Nature}} \textbf{\bibinfo{volume}{460}},
  \bibinfo{pages}{724--727} (\bibinfo{year}{2009}).

\bibitem{Verhagen2011}
\bibinfo{author}{Verhagen, E.}, \bibinfo{author}{Deleglise, S.},
  \bibinfo{author}{Weis, S.}, \bibinfo{author}{Schliesser, A.} \&
  \bibinfo{author}{Kippenberg, T.~J.}
\newblock \bibinfo{title}{Quantum-coherent coupling of a mechanical oscillator
  to an optical cavity mode}.
\newblock \emph{\bibinfo{journal}{Nature}} \textbf{\bibinfo{volume}{482}},
  \bibinfo{pages}{63--67} (\bibinfo{year}{2012}).

\bibitem{Taylor2011}
\bibinfo{author}{Taylor, J.~M.}, \bibinfo{author}{S\o{}rensen, A.~S.},
  \bibinfo{author}{Marcus, C.~M.} \& \bibinfo{author}{Polzik, E.~S.}
\newblock \bibinfo{title}{Laser cooling and optical detection of excitations in
  a $lc$ electrical circuit}.
\newblock \emph{\bibinfo{journal}{Physical Review Letters}}
  \textbf{\bibinfo{volume}{107}}, \bibinfo{pages}{273601}
  (\bibinfo{year}{2011}).

\bibitem{Regal2011}
\bibinfo{author}{Regal, C.~A.} \& \bibinfo{author}{Lehnert, K.~W.}
\newblock \bibinfo{title}{From cavity electromechanics to cavity
  optomechanics}.
\newblock \emph{\bibinfo{journal}{Journal of Physics: Conference Series}}
  \textbf{\bibinfo{volume}{264}}, \bibinfo{pages}{012025}
  (\bibinfo{year}{2011}).

\bibitem{Safavi2011}
\bibinfo{author}{Safavi-Neini, A.~H.} \& \bibinfo{author}{Painter, O.}
\newblock \bibinfo{title}{Proposal for an optomechanical traveling wave
  phonon-photon translator}.
\newblock \emph{\bibinfo{journal}{New Journal of Physics}}
  \textbf{\bibinfo{volume}{13}}, \bibinfo{pages}{013017}
  (\bibinfo{year}{2011}).

\bibitem{Wang2012a}
\bibinfo{author}{Wang, Y.-D.} \& \bibinfo{author}{Clerk, A.~A.}
\newblock \bibinfo{title}{Using interference for high fidelity quantum state
  transfer in optomechanics}.
\newblock \emph{\bibinfo{journal}{Physical Review Letters}}
  \textbf{\bibinfo{volume}{108}} (\bibinfo{year}{2012}).

\bibitem{McGee2013}
\bibinfo{author}{McGee, S.~A.}, \bibinfo{author}{Meiser, D.},
  \bibinfo{author}{Regal, C.~A.}, \bibinfo{author}{Lehnert, K.~W.} \&
  \bibinfo{author}{Holland, M.~J.}
\newblock \bibinfo{title}{Mechanical resonators for storage and transfer of
  electrical and optical quantum states}.
\newblock \emph{\bibinfo{journal}{Physical Review A}}
  \textbf{\bibinfo{volume}{87}} (\bibinfo{year}{2013}).

\bibitem{Rabl2010}
\bibinfo{author}{Rabl, P.} \emph{et~al.}
\newblock \bibinfo{title}{A quantum spin transducer based on
  nanoelectromechanical resonator arrays}.
\newblock \emph{\bibinfo{journal}{Nature Physics}}
  \textbf{\bibinfo{volume}{6}}, \bibinfo{pages}{602--608}
  (\bibinfo{year}{2010}).

\bibitem{Stannigel2010}
\bibinfo{author}{Stannigel, K.}, \bibinfo{author}{Rabl, P.},
  \bibinfo{author}{Sorensen, A.~S.}, \bibinfo{author}{Zoller, P.} \&
  \bibinfo{author}{Lukin, M.~D.}
\newblock \bibinfo{title}{Optomechanical transducers for long-distance quantum
  communication}.
\newblock \emph{\bibinfo{journal}{Physical Review Letters}}
  \textbf{\bibinfo{volume}{105}}, \bibinfo{pages}{220501}
  (\bibinfo{year}{2010}).

\bibitem{Dong2012}
\bibinfo{author}{Dong, C.}, \bibinfo{author}{Fiore, V.},
  \bibinfo{author}{Kuzyk, M.~C.} \& \bibinfo{author}{Wang, H.}
\newblock \bibinfo{title}{Optomechanical dark mode}.
\newblock \emph{\bibinfo{journal}{Science}} \textbf{\bibinfo{volume}{338}},
  \bibinfo{pages}{1609} (\bibinfo{year}{2012}).

\bibitem{Lee2010}
\bibinfo{author}{Lee, K.~H.}, \bibinfo{author}{McRae, T.~G.},
  \bibinfo{author}{Harris, G.~I.}, \bibinfo{author}{Knittel, J.} \&
  \bibinfo{author}{Bowen, W.~P.}
\newblock \bibinfo{title}{Cooling and control of a cavity
  opto-electromechanical system}.
\newblock \emph{\bibinfo{journal}{Physical Review Letters}}
  \textbf{\bibinfo{volume}{104}}, \bibinfo{pages}{123604}
  (\bibinfo{year}{2010}).

\bibitem{Thompson2007}
\bibinfo{author}{Thompson, J.~D.} \emph{et~al.}
\newblock \bibinfo{title}{Strong dispersive coupling of a high finesse cavity
  to a micromechanical membrane}.
\newblock \emph{\bibinfo{journal}{Nature}} \textbf{\bibinfo{volume}{452}},
  \bibinfo{pages}{72--75} (\bibinfo{year}{2008}).

\bibitem{Purdy2013}
\bibinfo{author}{Purdy, T.~P.}, \bibinfo{author}{Peterson, R.~W.} \&
  \bibinfo{author}{Regal, C.~A.}
\newblock \bibinfo{title}{Observation of radiation pressure shot noise on a
  macroscopic object}.
\newblock \emph{\bibinfo{journal}{Science}} \textbf{\bibinfo{volume}{339}},
  \bibinfo{pages}{801--804} (\bibinfo{year}{2013}).

\bibitem{Unterreithmeier2009}
\bibinfo{author}{Unterreithmeier, Q.~P.}, \bibinfo{author}{Weig, E.~M.} \&
  \bibinfo{author}{Kotthaus, J.~P.}
\newblock \bibinfo{title}{Universal transduction scheme for nanomechanical
  systems based on dielectric forces}.
\newblock \emph{\bibinfo{journal}{Nature}} \textbf{\bibinfo{volume}{458}},
  \bibinfo{pages}{1001--1004} (\bibinfo{year}{2009}).

\bibitem{Schmid2010}
\bibinfo{author}{Schmid, S.}, \bibinfo{author}{Hierold, C.} \&
  \bibinfo{author}{Boisen, A.}
\newblock \bibinfo{title}{Modeling the kelvin polarization force actuation of
  micro- and nanomechanical systems}.
\newblock \emph{\bibinfo{journal}{Journal of Applied Physics}}
  \textbf{\bibinfo{volume}{107}}, \bibinfo{pages}{054510}
  (\bibinfo{year}{2010}).

\bibitem{Faust2012}
\bibinfo{author}{Faust, T.}, \bibinfo{author}{Krenn, P.},
  \bibinfo{author}{Manus, S.}, \bibinfo{author}{Kotthaus, J.} \&
  \bibinfo{author}{Weig, E.}
\newblock \bibinfo{title}{Microwave cavity-enhanced transduction for plug and
  play nanomechanics at room temperature}.
\newblock \emph{\bibinfo{journal}{Nature Communications}}
  \textbf{\bibinfo{volume}{3}}, \bibinfo{pages}{728} (\bibinfo{year}{2012}).

\bibitem{Yu2012}
\bibinfo{author}{Yu, P.-L.}, \bibinfo{author}{Purdy, T.~P.} \&
  \bibinfo{author}{Regal, C.~A.}
\newblock \bibinfo{title}{Control of material damping in high-q membrane
  microresonators}.
\newblock \emph{\bibinfo{journal}{Physical Review Letters}}
  \textbf{\bibinfo{volume}{108}}, \bibinfo{pages}{083603}
  (\bibinfo{year}{2012}).

\bibitem{Schliesser2010}
\bibinfo{author}{Schliesser, A.} \& \bibinfo{author}{Kippenberg, T.~J.}
\newblock \bibinfo{title}{Cavity optomechanics with whispering-gallery-mode
  optical microresonators}.
\newblock In \bibinfo{editor}{Arimondo, E.}, \bibinfo{editor}{Berman, P.} \&
  \bibinfo{editor}{Lin, C.~C.} (eds.) \emph{\bibinfo{booktitle}{Advances in
  atomic, molecular and optical physics}}, vol.~\bibinfo{volume}{58},
  chap.~\bibinfo{chapter}{5}, \bibinfo{pages}{207--323}
  (\bibinfo{publisher}{Elsevier Academic Press}, \bibinfo{year}{2010}).

\bibitem{Weis2010}
\bibinfo{author}{Weis, S.} \emph{et~al.}
\newblock \bibinfo{title}{Optomechanically induced transparency}.
\newblock \emph{\bibinfo{journal}{Science}} \textbf{\bibinfo{volume}{330}},
  \bibinfo{pages}{1520--1523} (\bibinfo{year}{2010}).

\bibitem{Zhou2013}
\bibinfo{author}{Zhou, X.} \emph{et~al.}
\newblock \bibinfo{title}{Slowing, advancing and switching of microwave signals
  using circuit nanoelectromechanics}.
\newblock \emph{\bibinfo{journal}{Nature Physics}}
  \textbf{\bibinfo{volume}{9}}, \bibinfo{pages}{179--184}
  (\bibinfo{year}{2013}).

\bibitem{Gigan2006}
\bibinfo{author}{Gigan, S.} \emph{et~al.}
\newblock \bibinfo{title}{Self-cooling of a micromirror by radiation pressure}.
\newblock \emph{\bibinfo{journal}{Nature}} \textbf{\bibinfo{volume}{444}},
  \bibinfo{pages}{67--70} (\bibinfo{year}{2006}).

\bibitem{Arcizet2006a}
\bibinfo{author}{Arcizet, O.}, \bibinfo{author}{Cohadon, P.-F.},
  \bibinfo{author}{Briant, T.}, \bibinfo{author}{Pinard, M.} \&
  \bibinfo{author}{Heidmann, A.}
\newblock \bibinfo{title}{Radiation-pressure cooling and optomechanical
  instability of a micromirror}.
\newblock \emph{\bibinfo{journal}{Nature}} \textbf{\bibinfo{volume}{444}},
  \bibinfo{pages}{71--74} (\bibinfo{year}{2006}).

\bibitem{Schliesser2006}
\bibinfo{author}{Schliesser, A.}, \bibinfo{author}{Del'Haye, P.},
  \bibinfo{author}{Nooshi, N.}, \bibinfo{author}{Vahala, K.} \&
  \bibinfo{author}{Kippenberg, T.}
\newblock \bibinfo{title}{Radiation pressure cooling of a micromechanical
  oscillator using dynamical backaction}.
\newblock \emph{\bibinfo{journal}{Physical Review Letters}}
  \textbf{\bibinfo{volume}{97}}, \bibinfo{pages}{243905}
  (\bibinfo{year}{2006}).

\bibitem{Marquardt2007}
\bibinfo{author}{Marquardt, F.}, \bibinfo{author}{Chen, J.~P.},
  \bibinfo{author}{Clerk, A.~A.} \& \bibinfo{author}{Girvin, S.~M.}
\newblock \bibinfo{title}{Quantum theory of cavity-assisted sideband cooling of
  mechanical motion}.
\newblock \emph{\bibinfo{journal}{Physical Review Letters}}
  \textbf{\bibinfo{volume}{99}}, \bibinfo{pages}{093902}
  (\bibinfo{year}{2007}).

\bibitem{Dobrindt2008}
\bibinfo{author}{Dobrindt, J.~M.}, \bibinfo{author}{Wilson-Rae, I.} \&
  \bibinfo{author}{Kippenberg, T.~J.}
\newblock \bibinfo{title}{Parametric normal-mode splitting in cavity
  optomechanics}.
\newblock \emph{\bibinfo{journal}{Physical Review Letters}}
  \textbf{\bibinfo{volume}{101}}, \bibinfo{pages}{263602}
  (\bibinfo{year}{2008}).

\bibitem{Ilchenko2008b}
\bibinfo{author}{Ilchenko, V.~S.} \emph{et~al.}
\newblock \bibinfo{title}{$k_a$-band all-resonant photonic microwave receiver}.
\newblock \emph{\bibinfo{journal}{IEEE Photonics Technology Letters}}
  \textbf{\bibinfo{volume}{20}}, \bibinfo{pages}{1600--1612}
  (\bibinfo{year}{2008}).

\bibitem{Devgan2010}
\bibinfo{author}{Devgan, P.~S.}, \bibinfo{author}{Pruessner, M.~W.},
  \bibinfo{author}{Urick, V.~J.} \& \bibinfo{author}{Williams, K.~J.}
\newblock \bibinfo{title}{Detecting low-power and {RF} signals and using a
  multimode optoelectronic and oscillator and integrated optical filter}.
\newblock \emph{\bibinfo{journal}{IEEE Photonics Technology Letters}}
  \textbf{\bibinfo{volume}{22}}, \bibinfo{pages}{152--154}
  (\bibinfo{year}{2010}).

\bibitem{Hosseinzadeh2011}
\bibinfo{author}{Hossein-Zadeh, M.} \& \bibinfo{author}{Levi, A.}
\newblock \bibinfo{title}{Ring resonator-based photonic microwave receiver
  modulator with picowatt sensitivity}.
\newblock \emph{\bibinfo{journal}{IET Optoelectronics}}
  \textbf{\bibinfo{volume}{5}}, \bibinfo{pages}{36} (\bibinfo{year}{2011}).

\bibitem{Tsang2010}
\bibinfo{author}{Tsang, M.}
\newblock \bibinfo{title}{Cavity quantum electro-optics}.
\newblock \emph{\bibinfo{journal}{Physical Review A}}
  \textbf{\bibinfo{volume}{81}}, \bibinfo{pages}{063837}
  (\bibinfo{year}{2010}).

\bibitem{Kraus1966}
\bibinfo{author}{Kraus, J.~D.}
\newblock \emph{\bibinfo{title}{Radio Astronomy}} (\bibinfo{publisher}{McGraw},
  \bibinfo{year}{1966}).

\bibitem{Anetsberger2009}
\bibinfo{author}{Anetsberger, G.} \emph{et~al.}
\newblock \bibinfo{title}{Near-field cavity optomechanics with nanomechanical
  oscillators}.
\newblock \emph{\bibinfo{journal}{Nature Physics}}
  \textbf{\bibinfo{volume}{5}}, \bibinfo{pages}{909--914}
  (\bibinfo{year}{2009}).

\bibitem{Kovacs2005}
\bibinfo{author}{Kovacs, H.}, \bibinfo{author}{Moskau, D.} \&
  \bibinfo{author}{Spraul, M.}
\newblock \bibinfo{title}{Cryogenically cooled probes---a leap in {NMR}
  technology}.
\newblock \emph{\bibinfo{journal}{Progress in Nuclear Magnetic Resonance
  Spectroscopy}} \textbf{\bibinfo{volume}{46}}, \bibinfo{pages}{131--155}
  (\bibinfo{year}{2005}).

\end{thebibliography}

\begin{thebibliography}{1}
\expandafter\ifx\csname url\endcsname\relax
  \def\url#1{\texttt{#1}}\fi
\expandafter\ifx\csname urlprefix\endcsname\relax\def\urlprefix{URL }\fi
\providecommand{\bibinfo}[2]{#2}
\providecommand{\eprint}[2][]{\url{#2}}

\bibitem{SIKraus1966}
\bibinfo{author}{Kraus, J.~D.}
\newblock \emph{\bibinfo{title}{Radio Astronomy}} (\bibinfo{publisher}{McGraw},
  \bibinfo{year}{1966}).

\bibitem{SIChowdhury2005}
\bibinfo{author}{Chowdhury, S.}, \bibinfo{author}{Ahmadi, M.} \&
  \bibinfo{author}{Miller, W.~C.}
\newblock \bibinfo{title}{A closed-form model for the pull-in voltage of
  electrostatically actuated cantilever beams}.
\newblock \emph{\bibinfo{journal}{Journal of Micromechanics and
  Microengineering}} \textbf{\bibinfo{volume}{15}}, \bibinfo{pages}{756--763}
  (\bibinfo{year}{2005}).

\bibitem{SITaylor2011}
\bibinfo{author}{Taylor, J.~M.}, \bibinfo{author}{S\o{}rensen, A.~S.},
  \bibinfo{author}{Marcus, C.~M.} \& \bibinfo{author}{Polzik, E.~S.}
\newblock \bibinfo{title}{Laser cooling and optical detection of excitations in
  a $lc$ electrical circuit}.
\newblock \emph{\bibinfo{journal}{Physical Review Letters}}
  \textbf{\bibinfo{volume}{107}}, \bibinfo{pages}{273601}
  (\bibinfo{year}{2011}).

\bibitem{SIGroblacher2009a}
\bibinfo{author}{Gr{\"o}blacher, S.}, \bibinfo{author}{Hammerer, K.},
  \bibinfo{author}{Vanner, M.~R.} \& \bibinfo{author}{Aspelmeyer, M.}
\newblock \bibinfo{title}{Observation of strong coupling between a
  micromechanical resonator and an optical cavity field}.
\newblock \emph{\bibinfo{journal}{Nature}} \textbf{\bibinfo{volume}{460}},
  \bibinfo{pages}{724--727} (\bibinfo{year}{2009}).

\bibitem{SIVerhagen2011}
\bibinfo{author}{Verhagen, E.}, \bibinfo{author}{Deleglise, S.},
  \bibinfo{author}{Weis, S.}, \bibinfo{author}{Schliesser, A.} \&
  \bibinfo{author}{Kippenberg, T.~J.}
\newblock \bibinfo{title}{Quantum-coherent coupling of a mechanical oscillator
  to an optical cavity mode}.
\newblock \emph{\bibinfo{journal}{Nature}} \textbf{\bibinfo{volume}{482}},
  \bibinfo{pages}{63--67} (\bibinfo{year}{2012}).

\bibitem{SIPurdy2013}
\bibinfo{author}{Purdy, T.~P.}, \bibinfo{author}{Peterson, R.~W.} \&
  \bibinfo{author}{Regal, C.~A.}
\newblock \bibinfo{title}{Observation of radiation pressure shot noise on a
  macroscopic object}.
\newblock \emph{\bibinfo{journal}{Science}} \textbf{\bibinfo{volume}{339}},
  \bibinfo{pages}{801--804} (\bibinfo{year}{2013}).

\bibitem{SICouplingFrameworkInPrep}
\bibinfo{author}{Zeuthen, E.} \emph{et~al.}
\newblock \bibinfo{title}{In preparation} .

\end{thebibliography}
\end{document}